\documentclass[twocolumn,twocolappendix]{aastex701}   
 
\hypersetup{linkcolor=red,citecolor=blue,urlcolor=magenta}

\usepackage{newtxtext,newtxmath} 
 
\newcommand{\hunit}{km s$^{-1}$ Mpc$^{-1}$}

\newcommand{\kmps}{km s$^{-1}$}
\newcommand{\remaj}{\textit{$R_{\rm e}^{\rm maj}$}}
\newcommand{\as}{$^{\prime \prime}$}
\newcommand{\lm}{\textit{$\lambda_{R_\mathrm{e}}$}}
\newcommand{\lmr}{\textit{$\lambda_{R}$}}

\NewDocumentCommand\michele{O{Rewritten}m}{\textsf{\color{green}[Michele: #1]\(\rightarrow\)[``\emph{#2}'']}}

\begin{document}


\title{\LARGE  MAGNUS II: Rotational support of massive early-type galaxies decreased over the past 7 billion years}

\author[0000-0002-8593-7243]{Pritom Mozumdar}
\email{pmozumdar@astro.ucla.edu}
\affiliation{Department of Physics and Astronomy, University of California, Los Angeles, CA 90095, USA}
\affiliation{Department of Physics and Astronomy, University of California, Davis, 1 Shields Ave., Davis, CA 95616, USA}

\author[0000-0002-1283-8420]{Michele Cappellari}
\email{michele.cappellari@physics.ox.ac.uk}
\affiliation{Sub-Department of Astrophysics, Department of Physics, University of Oxford, Denys Wilkinson Building, Keble Road, Oxford, OX1 3RH, UK}


\author[0000-0002-4030-5461]{Christopher D. Fassnacht}
\email{fassnacht@ucdavis.edu}
\affiliation{Department of Physics and Astronomy, University of California, Davis, 1 Shields Ave., Davis, CA 95616, USA}

\author[0000-0002-8460-0390]{Tommaso Treu}
\email{tommaso.treu@physics.ucla.edu}
\affiliation{Department of Physics and Astronomy, University of California, Los Angeles, CA 90095, USA}

\correspondingauthor{Pritom Mozumdar}
\email{pmozumdar@astro.ucla.edu}

\begin{abstract}
Understanding how the internal kinematics of massive galaxies evolve is key to constraining the physical processes that drive their assembly. We investigate the evolution of rotational support in massive ($\log M_{\ast}/M_{\odot} \geq 10.6$) early-type galaxies (ETGs) over the past $\sim$7 Gyr. We use MUSE integral-field spectroscopic (IFS) data for 212 ETGs at intermediate redshift ($0.25 < z < 0.75$) from the MAGNUS sample. We compare their kinematics to a carefully matched local sample of 787 ETGs ($z \leq 0.05$) from the MaNGA survey. Using the specific stellar angular momentum proxy, $\lambda_R$, we quantify the balance between ordered rotation and random motions. We derive intrinsic $\lambda_R$ values by applying a uniform correction for seeing and point-spread function (PSF) effects to both samples. We find a significant evolutionary trend: the intermediate-redshift ETGs are systematically more rotationally supported than their local counterparts. The median PSF-corrected $\lambda_R$ for the MAGNUS sample is $0.48 \pm 0.05$, substantially higher than the median of $0.34 \pm 0.03$ for the matched MaNGA sample. This corresponds to a positive slope in the $\lambda_R-z$ relation of $\mathrm{d} \lambda_R / \mathrm{d} z = 0.25 \pm 0.04$ for the combined sample. The decline in rotational support is most pronounced for the most massive ($\log M_{\ast}/M_{\odot} > 11.3$) and highest-velocity-dispersion ($\sigma_e \gtrsim 200\,{\rm km\,s^{-1}}$) galaxies. Our results provide robust evidence that massive ETGs have undergone significant kinematic evolution, losing angular momentum as they evolve towards the present day, consistent with theoretical models where processes such as dry mergers play a crucial role in shaping the dynamical state of galaxies.
\end{abstract}


\section{Introduction} \label{sec:intro}


The dynamical state of galaxies reflects the balance between ordered rotation, typically quantified by the rotational velocity, and random stellar motions, measured via the velocity dispersion. The relative contributions of these processes determine whether a galaxy is rotation-supported or dispersion-supported. Understanding how this balance has changed across cosmic time offers critical insights into the physical processes that govern galaxy assembly and evolution \citep{Naab_2014, Lagos_2018b}. See \citet{Naab_Ostriker_2017} for a theoretical review. \\

With the advent of the integral field unit (IFU), a powerful tool to quantify this balance is the radially weighted dimensionless parameter $\lambda_R$, which combines information from galaxy light, line-of-sight stellar velocity, and velocity-dispersion fields \citep{Kinematic_classification_Emsellem_2007}. Hence, \lmr\ is robust against projection effects compared to simple $V/\sigma$ ratios, and is now widely adopted as the standard metric for kinematic classification \citep{ATLAS_III_Emselem_2011, Vande_sande_SAMI_2017}. It has enabled the division of early-type galaxies (ETGs) into fast and slow rotators, corresponding broadly to systems with significant rotational support and those dominated by random motions, respectively. See \citet{Cappellari_review_2016} for an observational review.\\

From theoretical perspectives, such as hydrodynamical simulations, a consensus has emerged that galaxies usually formed as rotationally supported systems at early times,  with angular momentum subsequently reduced through a combination of dissipative processes, quenching, and mergers \citep{Naab_2014, Lagos_2018a, Schulze_2018}. \citet{Lagos_2018a} using the EAGLE simulation \citep{Crain_2015_EAGLE, Schaye_2015_EAGLE} found that, while wet (or gas-rich) mergers can maintain or even enhance $\lambda_R$,  dry (or gas-poor) mergers are highly efficient at reducing $\lambda_R$, and the latter are becoming increasingly common than the former towards $z = 0$.  Using Magneticum Pathfinder simulations, \citet{Schulze_2018} concluded that at around $z=2$ the ETG population is dominated by fast rotator galaxies, usually having high \lmr, but gradually shifts to low \lmr\ population at later times, and dry mergers play a crucial role in this redistribution. Another way to decrease \lmr\ is the quenching process, which destroys ordered motions in the galaxy \citep{Hopkins_2009, Martig_2009}.\\

Observational studies have also provided important evidence for this picture. Studies at high redshift ($z\sim2$) found that quiescent galaxies retain significant rotational support \citep[e.g.][]{van_Dokkum_2015, Wisnioski_2015, Belli_2017, Slob_2025} in contrast with galaxies of similar mass in the local Universe. These observations suggest an apparent evolution in rotational support across cosmic time in line with the proposed scenarios from simulations. At intermediate redshift, the LEGA-C survey at $z\sim0.8$ \citep{LEGAC_survey_Wel_2016} revealed a higher $V/\sigma$ ratio than local galaxies from the CALIFA survey \citep{CALIFA_survey_2012}, providing statistically significant evidence for a decrease in rotational support toward the local Universe \citep{LEGAC_Bezanson_2018}. More recently, using IFU data,  \citet{Derkenne_2024} compared the MAGPI sample at $z\sim0.3$ \citep{MAGPI_survey_design_Foster_2021} to local samples from MaNGA \citep{MaNGA_survey_Bundy_2015} and found no significant difference in $\lambda_R$ at the population level; given the short lookback time probed by this sample ($\sim$3--4 Gyr), this result is consistent with only modest evolution having occurred during this epoch. Likewise, using a sample of galaxies spanning $0.1<z<0.8$ from various MUSE surveys, \citet{Munoz_Lopez_2024} found no clear trend of $\lambda_R$ with redshift, though their sample included relatively few massive ETGs. In this regard, the question of whether and how rotational support evolves from intermediate redshift to the present day remains worth revisiting with a statistically significant sample and high-quality IFU data spanning a long redshift baseline. \\

In the second paper of the MAGNUS (`Muse-deep Analysis of Galaxy kinematics and dyNamical evolUtion across redShift') series, we take advantage of a substantial sample of ETGs at intermediate redshifts ($0.2 < z < 0.75$) to revisit this question. This sample was collected using MUSE-DEEP IFU data which is comparable to local IFU surveys in depth and spatial resolution, enabling robust measurements of $\lambda_R$. By directly comparing this sample to local ETGs from MaNGA, we test for evolutionary trends in $\lambda_R$ and evaluate whether the massive ETG population indeed shows evidence of decrease in rotational support over the past $\sim$ 7 Gyrs. \\

This paper is organized as follows. In Sect. \ref{sec:muse_sample}, we briefly describe the MAGNUS sample and measurement of relevant data products pertinent to this analysis. In Section~\ref{sec:manga_sample}, we introduce the MaNGA survey and data products, detailing the methodology used to measure the galaxy properties of the MaNGA sample, and discuss how a MaNGA ETG sample was selected to statistically match the MAGNUS ETG sample. The seeing correction of the observed \lmr\ of the MAGNUS sample and its effect on both samples are discussed in Section~\ref{sec:psf correction}. We present the results in Section~\ref{sec:results} and discuss their implications and other considerations in Section~\ref{sec:discussion}. Finally, we conclude and summarize the work in Section~\ref{sec:conclusion}. Throughout this study, we assume a flat $\Lambda$CDM cosmology with $H_0 = 70.0$ \hunit and $\Omega_{\rm m} = 0.3$ when necessary.\\

\section{MAGNUS sample and data products}\label{sec:muse_sample}

The first paper of the MAGNUS series \citep[][hereafter MAGNUS I]{MAGNUS_I}, discusses extensively the sample selection and the kinematic and photometric data extraction process, and how these data were used to measure the global kinematic, morphological, and stellar population properties. In this work, we used these measured values for the MAGNUS sample. For convenience of the reader we provide a brief description of the sample and data products. 

\subsection{MAGNUS sample}
MUSE is an IFU spectrograph at the Very Large Telescope (VLT) operated by the European Southern Observatory (ESO). In wide-field mode, supported by adaptive optics (AO), it provides a spatial sampling of 0.2\as\ across a 1$^\prime \times$ 1$^\prime$ field of view (FOV) \citep{MUSE_Bacon_2010}. The instrument covers the wavelength range 4800–9300 \AA\ with a reciprocal dispersion of 1.25 \r{A}, delivering an average spectral resolution of $R \sim 2000$, which corresponds to an instrumental dispersion of $\sigma_{\text{inst}} \sim 65$ km$\mathrm{s}^{-1}$ near 5000 \AA. The MUSE-DEEP\footnote{\url{https://doi.org/10.18727/archive/42}} collection \citep{ESO2017} combines observations from single or multiple MUSE programs targeting the same sky regions into single IFU datacubes, resulting in very high effective exposure times. Utilizing this high-SNR data, MAGNUS I compiled a sample of around 200 ETGs from 35 datacubes within the redshift range of 0.25-0.75 with resolved stellar kinematics extending to several effective radii. \\

The sample predominantly consists of massive, luminous ETGs. Their integrated velocity dispersions, $\sigma_\mathrm{e}$, range from $\sim$100 to 300 \kmps\ with a median of 180 \kmps. In the HST F814W band, the surface brightness spans 16.62–20.34 mag/arcsec$^2$ with a median of 18.27 mag/arcsec$^2$. Galaxy sizes extend from $\remaj \sim 0.6$ to 10 kpc, with $\sim$80\% of systems between 1 and 4 kpc (median 2.2 kpc). Stellar masses cover $M_{\ast} \sim 1.3\times10^{10} - 10^{12} M_{\odot}$, with a median of $1.3\times10^{11} M_{\odot}$.\\

The MUSE-DEEP dataset was processed with the standard MUSE reduction pipeline\footnote{https://www.eso.org/sci/software/pipelines/muse} \citep{MUSE_pipeline_Weilbacher_2020}, yielding reduced 3D datacubes and corresponding variance cubes for each galaxy. To extract resolved kinematics, spaxels with SNR $\geq 3\sigma$ above the background were selected and subsequently Voronoi binned. The spectra within each bin were coadded to produce a single high-SNR spectrum. Kinematics in each bin were then measured in the rest-frame wavelength range 3780–4600 \r{A} using the penalized pixel fitting code\footnote{https://pypi.org/project/ppxf/} \textsc{pPXF} \citep{Cappellari_2017_ppxf, LEGAC_ppxf_Cappellari_2023}. Three stellar template libraries were employed — Indo-US \citep{Indo_US_lib}, MILES \citep{MILES_library_2011}, and X-shooter Spectral Library (XSL) \citep{Verro_2022} — restricted to the clean subset of stars identified by \citet{Knabel_Mozumdar_2025}. To minimize bin-to-bin scatter, we adopted a reduced template set rather than fitting all stars in a library. This set was constructed by first fitting the central spectrum of each galaxy with the full clean library and then retaining only those templates with non-zero weights in the fit. Following \citet{Mozumdar_Knabel_2025}, the mean velocity, velocity dispersion, and their associated systematic and statistical uncertainties for each bin were defined as the equally weighted average of the measurements obtained from the three libraries. Galaxy cutouts from HST imaging, primarily in the F814W filter (with F555W or F606W used in a few cases), were modeled with the \textsc{MgeFit} package\footnote{https://pypi.org/project/mgefit/} \citep{MGE_Cappellari_2002} to represent their surface brightness as a sum of 2D Gaussians.

\subsection{MAGNUS data products}
To probe the possibility of increased rotational support at intermediate redshift ETGs relative to local Universe ones, we used the following galaxy properties measured in MAGNUS I -- integrated velocity dispersion $\sigma_{\mathrm{e}}$, semi-major axis, \remaj, a proxy for stellar angular momentum \lmr, and stellar mass $M_{\ast}$. Here is a brief description of these properties and their measurement process -

(i) The integrated velocity dispersion, $\sigma_{\mathrm{e}}$, was measured from a central spectrum constructed by co-adding all spaxels within one effective radius ($R_\mathrm{e}$). Kinematics were extracted with the three stellar libraries described above, and $\sigma_{\mathrm{e}}$ was defined as the equally weighted average of the values obtained from these libraries.

(ii) The effective semi-major axis, \remaj, was derived from the corresponding MGE surface-brightness model of the galaxy within the elliptical isophote enclosing half of the total observed light. This measurement employed the `mge\_half\_light\_isophote' function within the \textsc{JamPy} package\footnote{https://pypi.org/project/jampy/} \citep{Jampy_Cappellari_2008}.

(iii) The stellar angular momentum parameter, \lmr, was computed following \citet{Kinematic_classification_Emsellem_2007} : 
\begin{equation}
\label{eq:lmr}
\lambda_R=\frac{\langle R|V|\rangle}{\left\langle R \sqrt{V^2+\sigma^2}\right\rangle}=\frac{\sum_{i=0}^N F_i R_i\left|V_i\right|}{\sum_{i=0}^N F_i R_i \sqrt{V_i^2+\sigma_i^2}},
\end{equation}
where $V_i$ is the stellar velocity, $\sigma_i$ is the velocity dispersion, $R_i$ is the distance from the galactic center, and $F_i$ is the flux of a spaxel within the adopted aperture. The values of $V_i$ and $\sigma_i$ were taken to be those of the Voronoi bin containing the spaxel. The summation was performed over all spaxels within an elliptical aperture whose semi-major axis was set to $2 \remaj$, with ellipticity and position angle fixed to those measured at the half-light isophote. For $\sim$10 galaxies, the aperture was expanded slightly to ensure a sufficient number of spaxels. Note that this is an observed quantity, and in Sect.~\ref{sec:psf correction}, we apply a correction for the instrument's point-spread-function (PSF) to derive its intrinsic value.

(iv) The stellar mass, $M_*$, was defined as the product of the total luminosity, $L$, and mass-to-light ratio, $M_*/L$. The total luminosity was measured from the MGE models in the observed HST band, accounting for cosmological dimming and extinction and applying a K-correction \citep{K-correction_Hogg_2002}. The $M_*/L$ values were derived from the central spectra using the \textsc{pPXF} package \citep{Cappellari_2017_ppxf} with FSPS stellar population synthesis templates \citep{FSPS_Conroy_2009, FSPS_Conroy_2010}.

\begin{figure*}
    \centering
    {\includegraphics[scale=0.5]{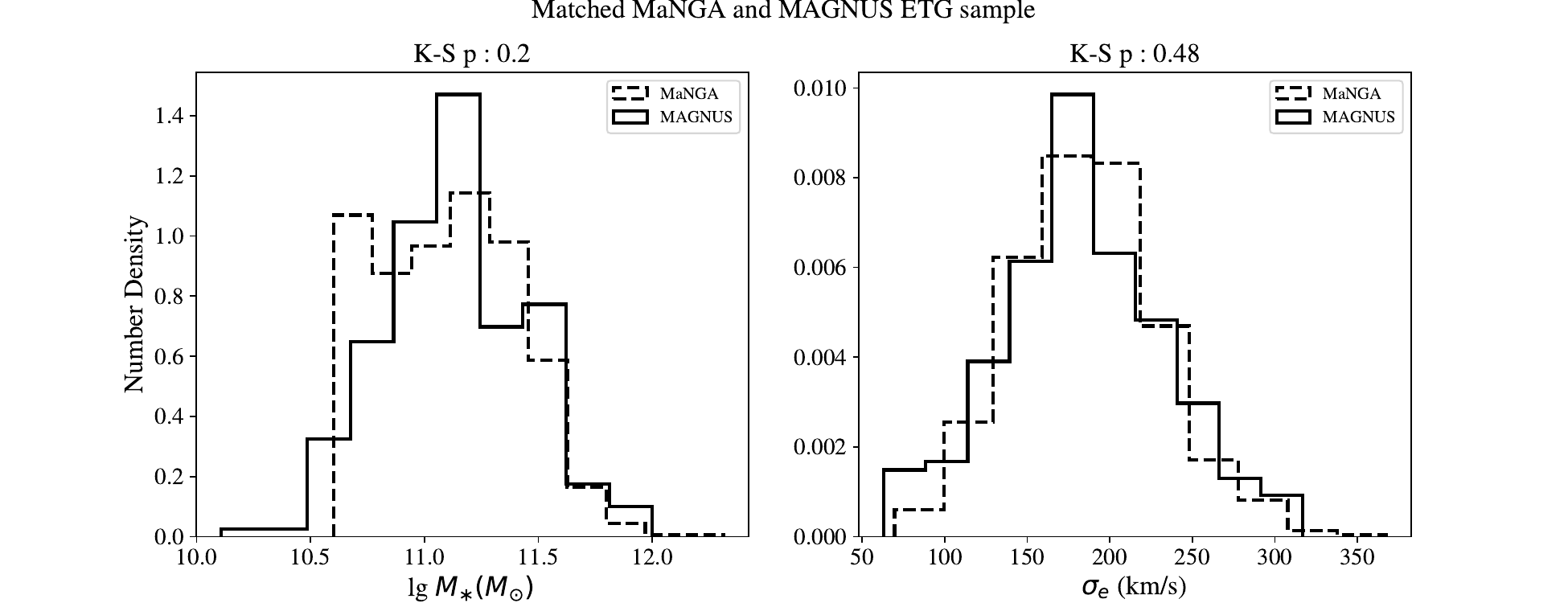}}
    \caption{Distribution in stellar mass $M_{\ast}$ (left) and stellar velocity dispersion $\sigma_{\mathrm{e}}$ (right) of the MaNGA ETG sample (dashed line) and MAGNUS sample (solid line). A mass cut of log $M_{\ast} (M_{\odot}) \geq 10.6$ and a redshift cut of $z \leq 0.05$ were applied to select the MaNGA ETG sample. 
    For both samples, the stellar mass was measured from SPS fitting using \textsc{pPXF} with templates from the FSPS model. 
    The p-values of the corresponding K-S tests are shown in the title of the respective plots.}
    \label{fig:matched_MaNGA}
\end{figure*}

\section{Reference MaNGA local sample}\label{sec:manga_sample}
As a local benchmark for testing the evolution of rotational support from z $\sim$ 0.5 to z $\sim$ 0, we use a sample of ETGs from the MaNGA survey \citep{MaNGA_survey_Bundy_2015}. In the following sections, we first provide an overview of the MaNGA survey and describe the methodology used to derive the relevant galaxy properties, and then explain how we selected a MaNGA ETG subsample matched to our MAGNUS sample.

\subsection{MaNGA survey}
The MaNGA survey offers spatially resolved spectral measurements for around 10000 nearby galaxies in the redshift range $0.01<z<0.15$. The observation covers a spatial range of up to 1.5 $R_e$ for the primary sample (66\% of total galaxies) and 2.5 $R_e$ for the secondary sample (33\% of the total galaxies) \citep{MaNGA_observation_Wake_2017}. The secondary sample is at a relatively higher redshift than the primary sample. The MaNGA spectra cover a wavelength range of $\lambda =$ 3600-10000 \r{A} with a spectral resolution of $R \approx 2000$ that corresponds to $\sigma_{\text{inst}}=72$ km $s^{-1}$ \citep{MaNGA_DRP_Law_2016}. The data cubes were extracted from spectrophotometrically calibrated raw data using the MaNGA data reduction pipeline \citep{MaNGA_DRP_Law_2016}. To measure stellar kinematic maps of the galaxies, the data cubes were Voronoi binned at a target SNR of 10 and fitted using the \textsc{pPXF} python package. These steps were performed with the official MaNGA data analysis pipeline \citep[DAP;][]{MaNGA_DAP_Westfall_2019}.

\begin{figure*}[ht]
    \centering
    {\includegraphics[width=\textwidth]{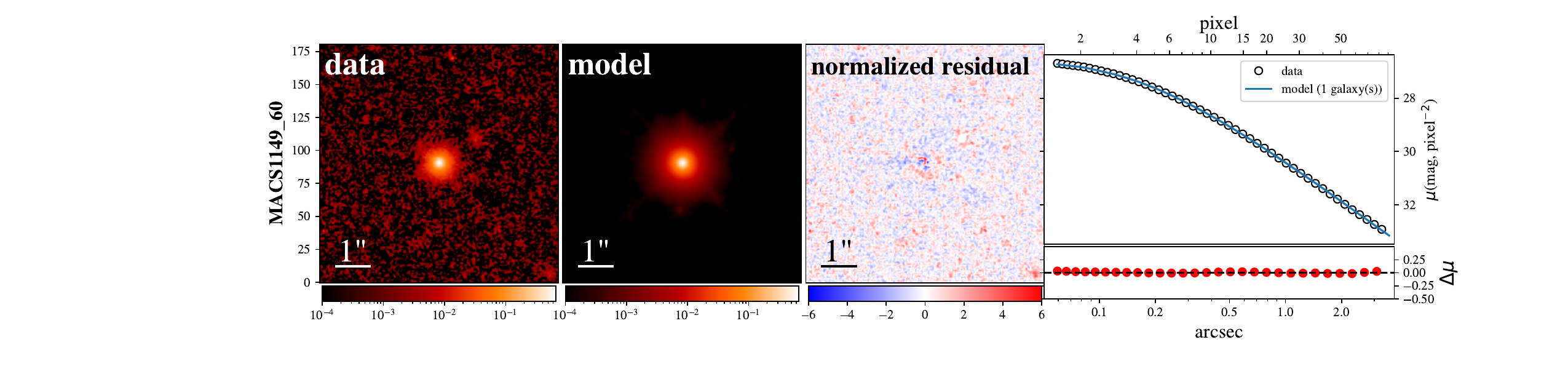}}
  \caption{S{\'e}rsic model fitting of a MAGNUS galaxy (MACS1149\_60, z=0.533) based on HST F814W image, using GALIGHT. The panels from left to right are as follows: (1) observed data, (2) best-fit model, (3) residuals divided by the variance, and (4) top-1D surface brightness profiles of data (open circles) and best-fit model (blue line); bottom - the corresponding residual. Note that the 1D surface brightness profiles are only for illustration purposes. The actual fitting is based on the 2D images.}
\label{fig:sersic fit}
\end{figure*}

\subsection{MaNGA data products} \label{sec:manga analysis}
The comprehensive dataset provided by the MaNGA survey has been extensively studied in the MaNGA Dynamics and Stellar Population (DynPop) series papers. The first of this series of papers provides a catalog of kinematic, structural, and dynamical properties for 10296 galaxies from the MaNGA survey \citep[][hereafter DynPop I]{Dynpop_I_Zhu_2023} and the second paper provides stellar population properties of this sample \citep[][hereafter DynPop II]{DynPop_II_Lu_2023}. For the MaNGA sample, we used the relevant galaxy properties from these two catalogs -- DynPop I catalog\footnote{\href{https://github.com/manga-dynpop/manga-dynpop.github.io/tree/main/catalogs/JAM}{https://github.com/manga-dynpop/manga-dynpop.github.io/tree/main/catalogs/JAM}} provides $\sigma_{e}$, PSF-corrected \lm, and \remaj, while DynPop II catalog\footnote{\href{https://doi.org/10.5281/zenodo.15742825}{https://doi.org/10.5281/zenodo.15742825}} \citep{DynPop_II_dataset} provides $M_{\ast}$. Here, we briefly describe the measurement process of the above-mentioned galaxy properties for the MaNGA sample:

(i) The integrated velocity dispersion $\sigma_{\mathrm{e}}$, was calculated as the square root of the luminosity-weighted average of the line-of-sight (LOS) velocity and velocity dispersion of the Voronoi bins within a circular aperture of radius $R_e$ such that --
\begin{equation}
    \sigma_{\mathrm{e}} = \sqrt{\frac{\sum_i F_i (V_i^2+\sigma_i^2)}{\sum_i F_i}}
\end{equation}
where $V_i$ and $\sigma_i$ are the velocity and velocity dispersion in the $i$ th Voronoi bin, respectively, and $F_i$ is the total flux in that bin. $R_e$ is the circularized effective radius of the elliptical half-light isophote. The $V_i$, $\sigma_i$, and $F_i$ are products of the MaNGA DAP.

(ii) The effective semi-major axis, \remaj, within the half-light isophote was measured similarly to that of the MAGNUS sample. The SDSS r-band images were used to create MGE models using \textsc{MgeFit}, which were then fitted with the `mge\_half\_light\_isophote' function from the \textsc{JamPy} package. 

(iii) The stellar angular momentum parameter, \lm, was calculated within an elliptical aperture of 1\remaj\ following equation \ref{eq:lmr}. The observed values were then corrected for PSF effects following equation 5 of \citet{Graham_2018}. The correction depends on the aperture size, PSF of the instrument, and S\'ersic index of the galaxy. The S\'ersic index values are available in DynPop I, and the SDSS r-band PSF FWHM values are available in a catalogue\footnote{\href{https://www.sdss.org/dr17/manga/manga-data/data-access/}{https://www.sdss.org/dr17/manga/manga-data/data-access/}} provided by the MaNGA collaboration. The DynPop I catalog provides the PSF-corrected \lm, and we collected the corresponding observed values from the DynPop I team through private communication. We applied a similar treatment to the MAGNUS sample in Sect. \ref{sec:psf correction}.

(iv) The stellar mass $M_{\ast}$ was measured following 
\begin{equation}
    M_{\ast} = (M_{\ast}/L)_{SPS}(<R_e) \times L
\end{equation}
where $(M_{\ast}/L)_{SPS}(<R_e)$ is the SDSS r-band stellar mass-to-light ratio derived by fitting an integrated spectrum from all spaxels within the elliptical half-light isophote, and using the \texttt{mass\_to\_light} procedure of the \textsc{ppxf} package, and $L$ is the r-band total luminosity calculated from the MGE model. The MGE surface brightness was converted to the AB magnitude system \citep{Oke1983} and corrected for cosmological dimming by multiplying the surface-brightness values by $(1+z)^3$, which accounts for the combined effects of cosmological surface-brightness dimming and the redshifting of the observed bandpass. However, the K-correction was not applied as most of the MaNGA galaxies are located at low redshift, and it is therefore negligible. \\

The SPS fitting was performed using the \textsc{pPXF} package. Stellar mass-to-light ratios were derived using one of the default SPS template set distributed with \textsc{pPXF}. These template sets were produced by the \textsc{fsps} stellar population models assuming a Salpeter IMF \citep{Salpeter1955} with lower and upper mass limits of 0.08 and 100 $M_{\odot}$, and employ the MIST isochrones \citep{MESA_isochrone_Choi_2016}. These templates span 43 ages, logarithmically spaced by 0.1 dex from 1 Myr to 15.85 Gyr (log Age/yr = 6.0, 6.1, …, 10.2), and nine metallicities [Z/H] = [-1.75, -1.50, -1.25, -1.00, -0.75, -0.50, -0.25, 0.00, +0.25], yielding $43\times9=387$ SPS templates in total.\\


In general, the measurement of the relevant galaxy properties for both the MAGNUS and MaNGA samples follows the same procedure and employs identical tools and techniques. Though $\sigma_e$ is measured slightly differently, this does not affect the analysis since it is used solely to define a MaNGA subsample. For \lmr, we adopted a 2 \remaj\ elliptical aperture, in contrast to the 1 \remaj\ elliptical aperture used in MaNGA. The impact of aperture size, S\'ersic index, and PSF on our results is discussed in Sect.~\ref{sec:results}.

\subsection{MaNGA sample selection}
DynPop I graded the MaNGA sample according to the quality of the dynamical model based on the 2D kinematics map as  -1, 0, 1, 2, 3 (from worst to best). In this work, we considered only those galaxies with $\text{Qual}\geq 0$ because, according to DynPop I, no kinematic measurements are reliable with $\text{Qual} = -1$. The MaNGA sample consists of elliptical (E), lenticular (S0), and spiral (S) galaxies. We obtained the morphological classification of these galaxies from the MaNGA Deep Learning (DL) morphological catalog \citep{MaNGA_morphology_catalog_Dominguez_2017}. Applying the prescribed morphological selection criteria recommended in \citet{MaNGA_morphology_catalog_Dominguez_2017}, and the above quality constraint, we found 2183 elliptical and 807 lenticular galaxies among 10128 unique MaNGA galaxies.

However, approximately 51\% of the MaNGA ETGs are located at $z > 0.05$. To construct a sample that truly represents the low-redshift Universe, we restricted the selection of `E' and `S0' type galaxies at $z \leq 0.05$. In addition, we required a stellar mass of $\log M_{\ast}(M_{\odot}) \geq 10.6$, corresponding to the lower bound of the stellar mass range of the MAGNUS sample. We did not apply an upper mass cut, since the two samples naturally overlap at the high-mass end. After applying these criteria, the final MaNGA ETG sample consists of 787 galaxies, of which 591 are classified as `E' and the remainder as `S0'. To test whether the MaNGA and MAGNUS ETG samples are consistent with being drawn from the same underlying distributions in $\sigma_e$ and $M_{\ast}$, we performed two-sample Kolmogorov–Smirnov (K–S) tests. The resulting $p$-values are 0.20 and 0.48, respectively, indicating statistical consistency. Figure~\ref{fig:matched_MaNGA} further illustrates this agreement, showing overlapping distributions of both $\log M_{\ast}$ and $\sigma_e$ for the two samples.

\section{Applying correction for PSF effects} \label{sec:psf correction}
The apparent sizes of galaxies and spatial resolution decrease with distance, making atmospheric seeing effects significant. It is essential to account for these effects, especially for quantities derived from line-of-sight (LOS) kinematic maps such as \lmr. \citet{Vande_sande_SAMI_2017} found that the observed \lmr\ systematically decreases with increasing PSF size, with the impact being strongest for compact galaxies where $R_e$ is comparable to the PSF. Moreover, galaxies with intrinsically higher \lmr\ are more strongly affected than those with lower values. For the MaNGA sample, the \lm\ measurements from DynPop I were already corrected for PSF effects. To ensure a fair comparison, we likewise applied PSF corrections to the measured \lmr\ of the MAGNUS galaxies. In the following sections, we describe the correction method, its application to the MAGNUS sample, and assess the impact of the correction on both the MAGNUS and MaNGA samples. 

\subsection{Methodology}

We followed the same analytic correction recipe used in DynPop~I. This procedure was developed by \citet{Graham_2018} using realistic Jeans Anisotropic Models (JAM) \citep{Jampy_Cappellari_2008} of galaxy kinematics and depends on the S{\'e}rsic index and on the ratio of the $\sigma_{\text{PSF}}$ to the effective semi-major axis, \remaj, of the galaxy. However, since in our analysis we measured \lmr\ within an aperture twice the effective semi-major axis, we assumed that the same correction formula derived by \citet{Graham_2018} for measurements within one \remaj\ also applies in this larger aperture. Therefore, following the same method, we applied the correction to the observed \lmr\ as defined in \citep[eqs.~5--7]{Graham_2018}. 


\begin{flalign}
\lambda_R^{\rm obs} = \lambda_R^{\rm corr} \, gM_2\left(r \right) f_n\left(r \right),
\end{flalign}
where
\begin{flalign}
r &= \frac{\sigma_{\rm PSF}}{R^{\rm maj}}, &&\\
gM_2\left(r \right) &= \left[ 1 + \left( \frac{r}{0.47} \right)^{1.76} \right]^{-0.84}, &&\\
f_n\left(r \right)  &= \left[ 1 + (n-2) \left( 0.26 r \right) \right]^{-1}&&
\end{flalign}
where $\lambda_R^{\rm obs}$ is the observed and $\lambda_R^{\rm corr}$ is the PSF-corrected angular momentum parameter measured in an elliptical aperture, $R^{\rm maj}$ is the semi-major axis of the aperture within which \lmr\ is calculated, and $n$ is the S{\'e}rsic index. The function $gM_2$ is a generalized form of the Moffat function \citep{Moffat_1969}, and $f_n$ is an empirical function used to model the dependence of the correction on the S{\'e}rsic index. However, a correction was only applied to a galaxy if the S{\'e}rsic index, $n$, was within the range of $0.5<n<6.5$ and the `resolution parameter', $r=\frac{\sigma_{\rm PSF}}{R^{\rm maj}} \leq 1$. Additionally, if the corrected \lmr\ value goes above 1, then the intrinsic \lmr\ of that galaxy was kept to its observed value. The error on $\lambda_R^{\rm corr}$ is calculated as $\left[ -0.08\,n r,  +0.03 r \right]$.

\begin{figure}
    \centering
    {\includegraphics[width=1.05\linewidth]{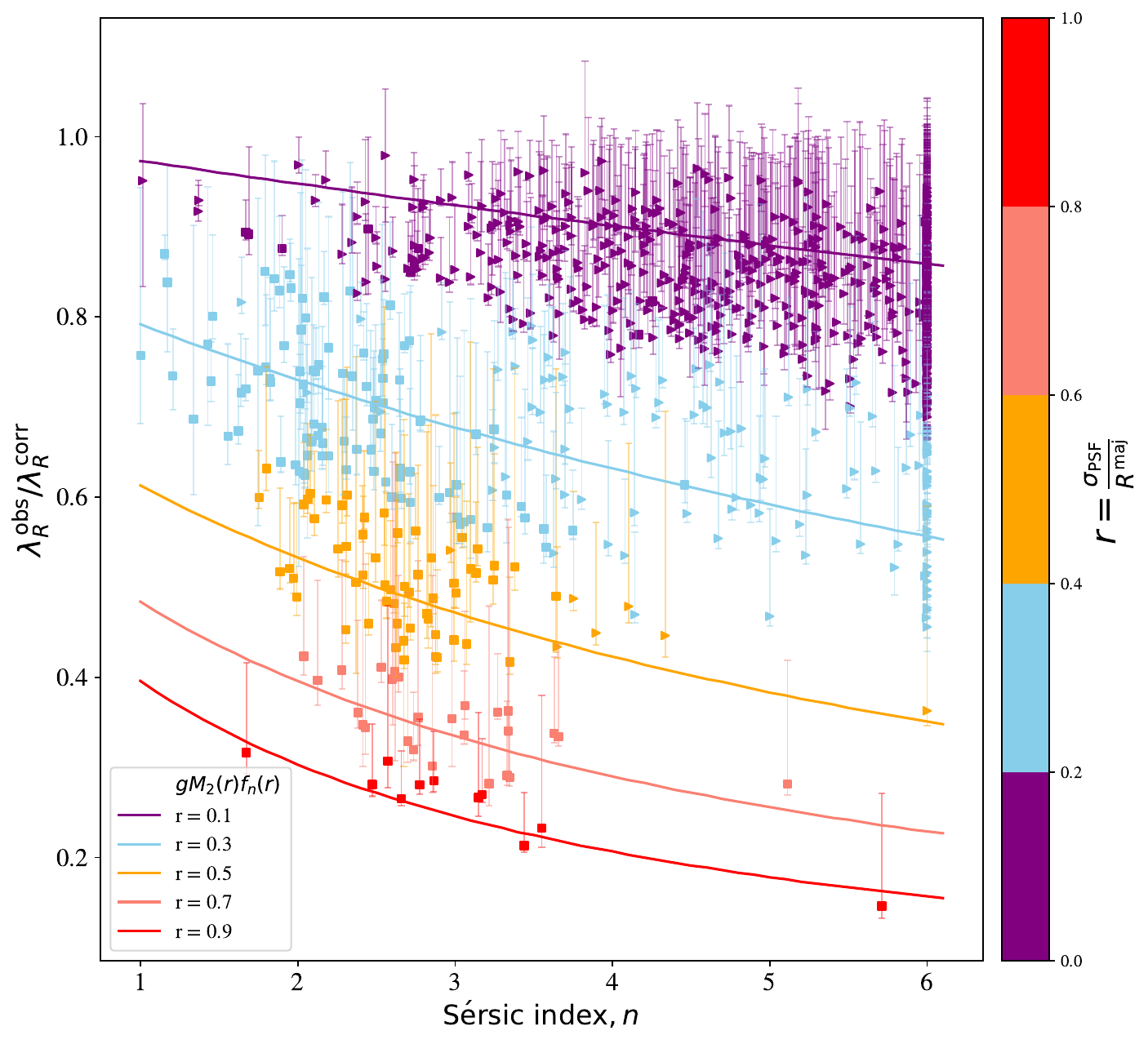}}
    \caption{Correction of the observed angular momentum parameter $\lambda_R^{\rm obs}$ for both samples. The ratio of observed to corrected \lmr\ is plotted as a function of the galaxy S{\'e}rsic index, $n$, and colored by the `resolution parameter' $r=\frac{\sigma_{\rm PSF}}{R^{\rm maj}}$, which quantifies how well resolved a galaxy is. For the MaNGA sample (right pointing triangle) $R^{\rm maj} = $ \remaj\ and for the MAGNUS galaxies (square) this is $R^{\rm maj} \approx $ 2\remaj. Solid lines show the effect of the S{\'e}rsic index on the corrections for a fixed value of $r$.}
    \label{fig:n_r_muse_manga}
\end{figure}

\begin{figure*}[ht]
    \centering
    {\includegraphics[width=\textwidth]{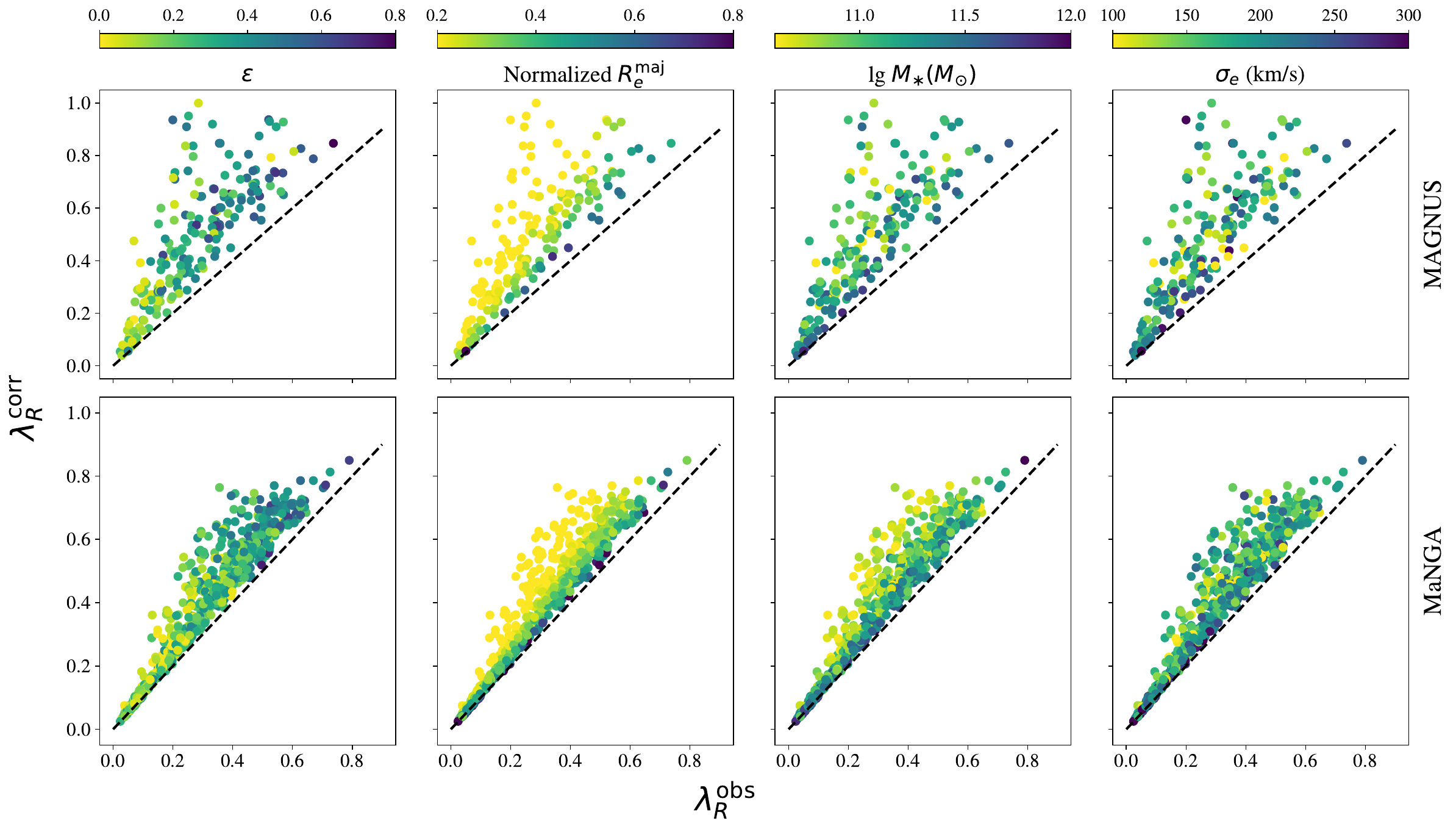}}
    \caption{Comparison of observed and PSF-corrected \lmr\ values for the MAGNUS (top row) and MaNGA sample (bottom row). From left to right, the values are colored with ellipticity, normalized semi-major axis, \remaj, stellar mass, $M_{\ast}$, and integrated velocity dispersion, $\sigma_e$. The black dashed line is the one-to-one line.}
    \label{fig:psf correction}
\end{figure*}

\subsection{PSF corrections for \lmr\ of MAGNUS sample}
For the MAGNUS sample, $R^{\rm maj} \sim$ 2\remaj, and the PSF (FWHM) values were collected from the FITS file of the respective MUSE-DEEP data cubes. Rather than adopting a global seeing value for the full FOV of a datacube , we used a position-specific estimate evaluated at the location of each galaxy. Because MUSE-DEEP datacubes are co-additions of observations spanning multiple programs, epochs, and atmospheric conditions, with the contributing exposures not necessarily covering the full FOV uniformly, the effective seeing varies across the combined cube; the position-specific estimate more accurately characterizes the local image quality at each galaxy's location. These estimates are provided in the header under the keywords \texttt{HIERARCH ESO QC EXPCOMB FWHM<N> X} and \texttt{HIERARCH ESO QC EXPCOMB FWHM<N> Y}, evaluated at the position recorded under \texttt{HIERARCH ESO QC EXPCOMB POS<N> X/Y} for each detected object \texttt{N}. We adopted the average of the two directional ($x$, $y$) estimates as a single circular PSF FWHM for each galaxy. For a small number of MAGNUS galaxies without an available position-specific estimate, we used the global seeing value (\texttt{'SKY\_RES'}) of the corresponding datacube instead. As an independent check on the reliability of these pipeline-based estimates, we measured the point-spread function directly from field stars in a subset of the MUSE-DEEP datacubes and compared the results to the corresponding pipeline provided values. This comparison is presented in Appendix~\ref{sec:appendix_psf}. We find that the pipeline values are systematically higher than the directly measured ones by a few percent, and we show that this offset has a negligible effect on our results. 

The S{\'e}rsic index of the MAGNUS galaxies was measured from their HST images using the \textsc{GaLight} package \citep{GaLight_Xuheng_2021}, which is based on the lens modeling code \textsc{lenstronomy} \citep{Lenstronomy_II_2021}. The software first detects the target galaxy along with neighboring objects, which are masked during the fitting process. A noise map is then constructed from the background regions of the image, and the PSF is modeled from stars within the field of view. The surface brightness distribution of each galaxy is then fitted with a single S{\'e}rsic profile, with parameter estimation carried out through a Markov Chain Monte Carlo (MCMC) Bayesian framework. In Fig.~\ref{fig:sersic fit}, we show an example S{\'e}rsic profile fit. The correction was not applied to 13 MAGNUS galaxies due to $\frac{\sigma_{\rm PSF}}{R^{\rm maj}} > 1$ or $\lambda_R^{\rm corr} > 1$. We excluded those galaxies from further analysis and got a final sample of 200 galaxies with corrected \lmr. The median error on $\lambda_R^{\rm corr}$ for the MAGNUS sample is 0.08 on the negative side and 0.01 on the positive side, and for the MaNGA sample, the corresponding values are 0.04 and 0.004. 


\begin{figure*}
    \centering
    {\includegraphics[width=\textwidth]{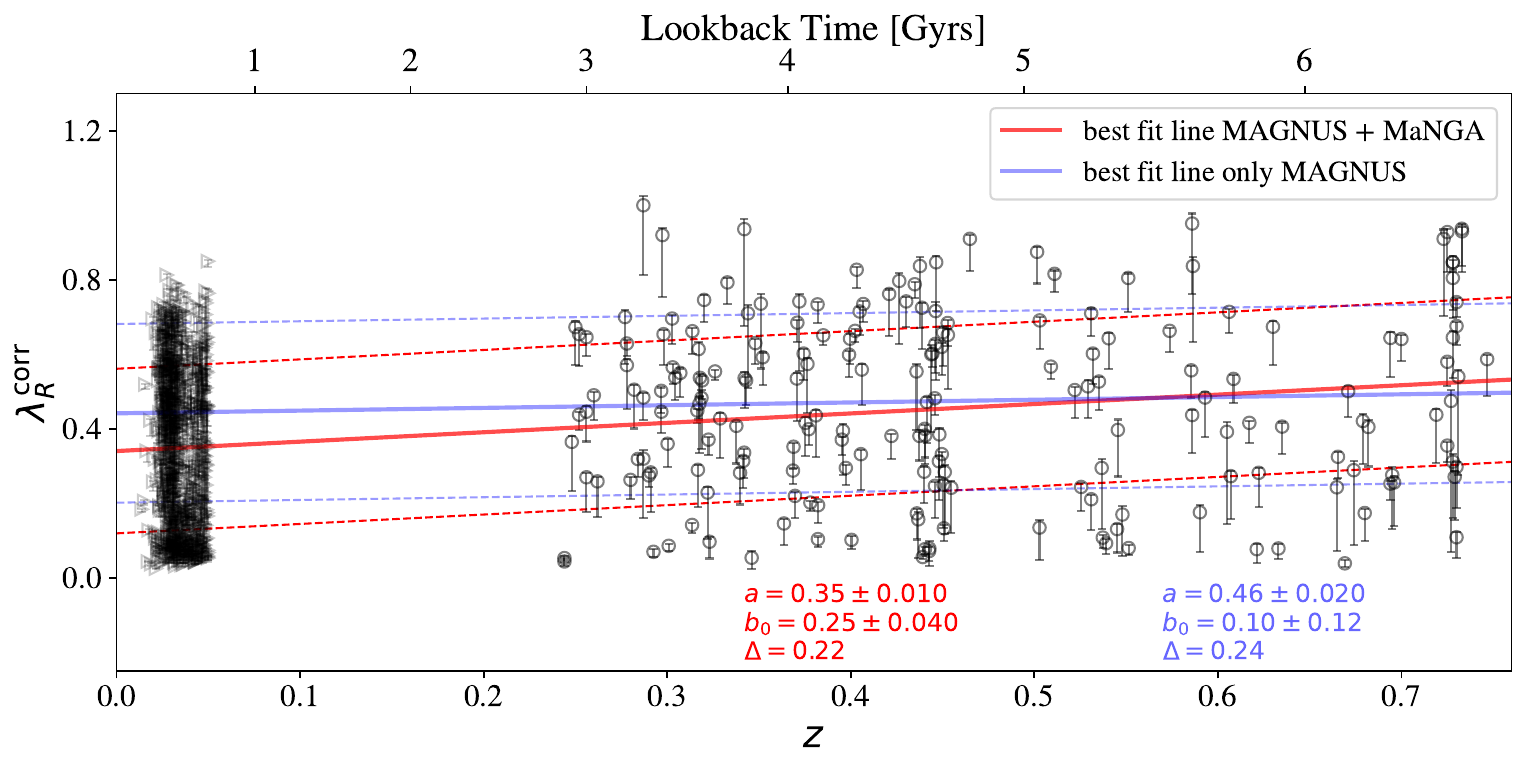}}
    \caption{PSF-corrected \lmr\ of both samples with redshift. The MAGNUS galaxies are marked using black circles, while the MaNGA sample is marked with black right-facing triangles. The solid lines show the best-fit line -- the blue one is from a linear fit with the MAGNUS sample alone, while the red one is from the combined sample. The dashed lines show the corresponding 1$\sigma$ uncertainty of the best-fit line. The parameters from the best-fit lines and associated intrinsic scatters are also reported using the respective color of the lines.}
    \label{fig:MaNGA_MUSE_lmr_z}
\end{figure*}

\subsection{Effect of correction on MAGNUS and MaNGA}
Since we have both observed and corrected \lmr\ values for MAGNUS and MaNGA, we next evaluate the impact of the correction on the two samples. Figure~\ref{fig:n_r_muse_manga} shows the ratio of observed to corrected \lmr\ for both samples as well as the role of the Sérsic index, $n$ and the `resolution parameter', $r$. From the figure, it is evident that $r$ is the dominant factor driving the correction. As the PSF smooths out light from the central region, the correction compensates for this effect. When $r$ is small (i.e., the aperture is large compared to the PSF width), the PSF effect is negligible, and the observed and corrected \lmr\ values are nearly identical. Conversely, at larger $r$, the PSF effect becomes more severe, requires a stronger correction, and leads to a smaller observed-to-corrected \lmr\ ratio. For $r > 1$, the correction becomes unreliable. The solid lines in Fig.~\ref{fig:n_r_muse_manga} illustrate how the Sérsic index modifies this trend: galaxies with a high Sérsic index ($n > 4$) have more centrally concentrated light and therefore require stronger corrections. The combined effect is that at poor resolution (large $r$), the correction is more affected by the surface brightness distribution (parametrized by $n$). \citet{Munoz_Lopez_2024} showed that for each $n$, there exists a maximum $r$ beyond which the correction becomes unphysical, with $\lambda_R^{\rm corr} > 1$. \\

As expected, the 
corrected \lmr\ values are systematically higher than their observed counterparts \citep{Graham_2018}. The effect is, however, much stronger for the MAGNUS sample than for MaNGA. For example, only $\sim$2\% of MaNGA galaxies have an observed-to-corrected \lmr\ ratio of $\leq$ 0.5 (i.e., their corrected value is twice the observed value), while this fraction rises to $\sim$30\% for MAGNUS galaxies. This difference arises because for the higher-redshift MAGNUS galaxies the ratio between angular size and seeing is smaller than in their low-redshift  MaNGA counterparts. By using a 2\remaj\ aperture for MAGNUS galaxies, instead of the 1\remaj\ aperture adopted for MaNGA, we minimize the influence of PSF effects on the measured \lmr. For MAGNUS galaxies with large $r$, this aperture represents the largest feasible measurement radius, while for others, the observed \lmr\ does not change significantly beyond this scale. Thus, the stronger corrections in MAGNUS are primarily resolution-driven, whereas in MaNGA they are more often associated with high Sérsic indices. The uncertainty in the observed-to-corrected \lmr\ ratio is calculated using standard error propagation, explicitly accounting for the asymmetric nature of the errors on $\lambda_R^{\rm corr}$.\\

In Fig.~\ref{fig:psf correction}, we compare observed and corrected \lmr\ values for both samples, color-coded by several galaxy properties including ellipticity, \remaj, $M_{\ast}$, and $\sigma_e$, to check for possible correlations with the correction. We find that for both samples, galaxies with $\lambda_R^{\rm obs} < 0.2$ are least affected, regardless of size or mass. At higher $\lambda_R^{\rm obs}$, both size and mass show some correlation with the strength of the correction. In particular, the most massive galaxies are generally less affected, irrespective of their observed \lmr, as they also tend to be larger and thus have lower $r$. The strong dependence on \remaj\ is clearly visible in both samples: smaller galaxies are more strongly impacted by the PSF correction, while larger galaxies are less so. Overall, both MAGNUS and MaNGA exhibit the same qualitative trends in how the correction depends (or not) on galaxy properties.


\begin{figure*}
    \centering
    {\includegraphics[width=\textwidth]{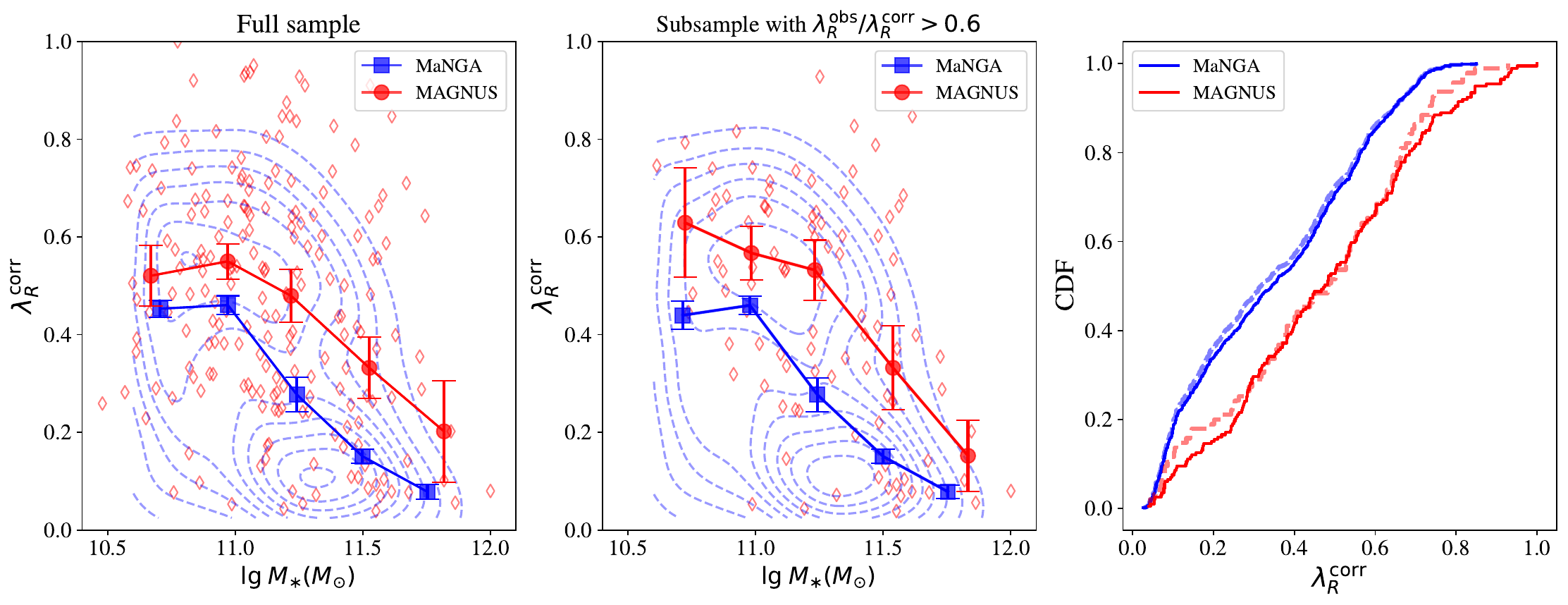}}
    \caption{Evidence of increased rotational support in high-redshift ETGs. Left: diamond shapes (light red) mark $\lambda_R^{\rm corr}$ values of the MAGNUS sample. The corresponding values from MaNGA galaxies are shown as a kernel density estimate (KDE) of the galaxy number density using light blue contour lines. The red circles denote the medians in bins of MAGNUS galaxies, and the blue squares are for similar quantities from MaNGA galaxies. Each set of medians and uncertainties was measured in bins of stellar mass with a width of 0.3 dex, with the first and last bins widened to include all galaxies beyond the fixed-width range, ensuring a sufficient number of galaxies at the extremes of the distribution. Middle: description of the plot is the same as the left panel, but this time the MAGNUS and MaNGA sample consists of galaxies satisfying  $\lambda_R^{\mathrm{obs}}/ \lambda_R^{\mathrm{corr}} > 0.6$. Right: CDF of the $\lambda_R^{\mathrm{corr}}$ distribution for full (solid lines) and sub-samples (dashed lines). The CDFs for MAGNUS and MaNGA are colored red and blue, respectively.}
    \label{fig:lr_mass}
\end{figure*}

\begin{figure*}
    \centering
    {\includegraphics[width=0.7\textwidth]{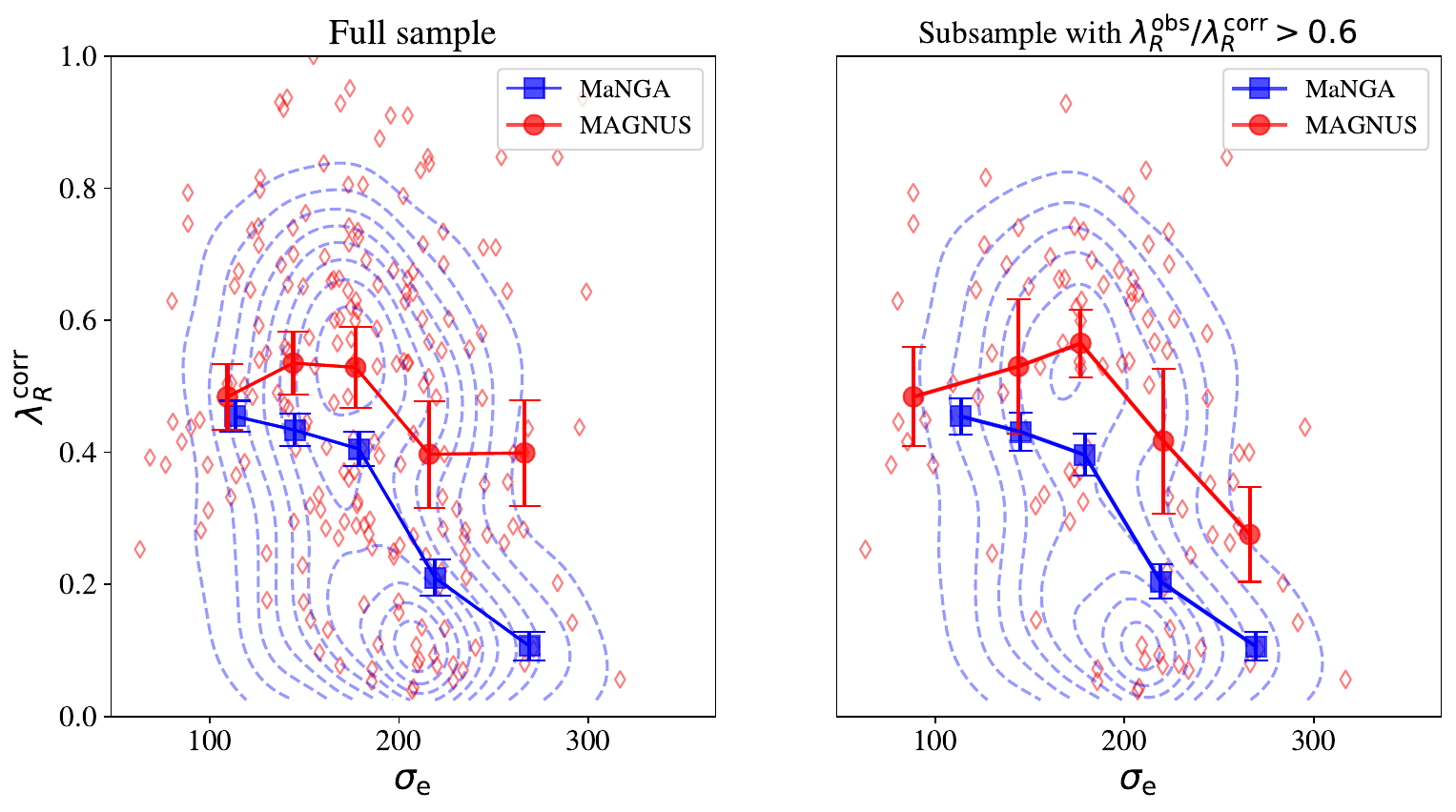}}
    \caption{Evidence of increased rotational support in high-redshift ETGs, binned by velocity dispersion. The panels follow the same description as in Fig.~\ref{fig:lr_mass} (left and middle panels), but with galaxies binned in $ \rm{log} \sigma_e$ with a bin width of $0.1$ dex instead of stellar mass.}
\label{fig:sigma_e_binned}
\end{figure*}

\section{Results}\label{sec:results}
With the PSF-corrected \lmr\ for the MAGNUS sample, we can evaluate the redshift evolution of rotational support towards low redshift by comparing this distribution with the matched MaNGA sample. The median value of $\lambda_R^{\rm corr}$ in the MAGNUS sample is $0.48 \pm 0.05$ while for the MaNGA sample it is $0.34 \pm 0.03$ and they originate from statistically distinguished populations. A two-sample K-S test also suggests that these two samples originate from statistically different populations in terms of $\lambda_R^{\rm corr}$ with a p-value of $7 \times 10^{-7}$. In Fig.~\ref{fig:MaNGA_MUSE_lmr_z}, we plot $\lambda_R^{\rm corr}$ of both samples as a function of redshift. We measured the redshift dependence of \lmr\ by conducting a linear fit using the \textsc{LtsFit} package \citep{ATLAS_XV_Michele_2013} and found that a fit to the combined MAGNUS and MaNGA sample produces a positive slope of $\mathrm{d} \lambda_R^{\rm corr}/ {\mathrm{d} z} = 0.25 \pm 0.04$ with a intrinsic scatter of 0.22. Even a fit to the MAGNUS galaxies alone yields a mild positive slope of $\mathrm{d} \lambda_R^{\rm corr}/ {\mathrm{d} z} = 0.1 \pm 0.12$ with an intrinsic scatter of 0.24. The difference arises because the combined dataset spans a wider redshift baseline, thus more strongly highlighting the evolution across cosmic time. In both cases, however, the results suggest a gradual decrease of rotational support from intermediate redshift ($z < 1$) to today.\\

Next, we examine whether the evidence of evolution in $\lambda_R^{\rm corr}$ only persists in particular sub-groups of galaxies. To evaluate this aspect, we allocated galaxies of both samples in bins of width 0.3 dex according to their stellar mass, with the first and last bins widened to include all galaxies beyond the fixed-width range, ensuring a sufficient number of galaxies at the extremes of the distribution. The number of galaxies in each bin for the MaNGA sample was, on average, four times higher than the MAGNUS sample. We then calculated medians and associated bootstrap uncertainties of $\lambda_R^{\rm corr}$ in each bin. In the leftmost plot of Fig.~\ref{fig:lr_mass}, we present the medians of each bin from both samples as a function of stellar mass. We found that across nearly the entire mass range, the MAGNUS galaxies display systematically higher median (on average $\sim$ 75\%) in $\lambda_R^{\rm corr}$ values than MaNGA galaxies. However, the difference is more prominent in relatively high mass ETGs ($\mathrm{lg}$ $M_{\ast} > 11.3 M_{\odot}$) than the low-mass ones. 
Varying the width of bins and thus the number of galaxies in each bin does not change the qualitative conclusion of this analysis. In the rightmost plot of Fig.~\ref{fig:lr_mass}, we show the cumulative distribution functions (CDFs) for the two samples, and except at the very low $\lambda_R^{\rm corr}$, the distributions remain distinctively different. A two-sample Anderson-Darling (AD) test, which is particularly sensitive to the tails of CDFs, also suggests that the samples are drawn from statistically distinguished populations with a p-value of 0.001.\\


To assess the robustness of these findings, we repeated the analysis on subsamples restricted to ETGs that are relatively less affected by the PSF corrections, thus minimizing possible bias from the correction. For this, we select sub-samples of ETGs from respective full samples with $\lambda_R^{\mathrm{obs}}/ \lambda_R^{\mathrm{corr}} > 0.6$. After applying this cut, the MaNGA subsample still retains 93\% ETGs (736 out of 787) while the MAGNUS subsample only retains 48\% galaxies (95 out of 200). However, the CDFs of the sub-samples remain essentially unchanged compared to the respective full samples, suggesting the PSF correction does not disproportionately affect any mass group, especially the MAGNUS sample. A two-sample AD test also confirms that the full and sub-MAGNUS samples are statistically indistinguishable with a p-value of 0.25. The medians, in the same mass bins used before, for these subsamples (middle plot in Fig.~\ref{fig:lr_mass}) exhibit the same behavior as the full samples, demonstrating the observed differences in $\lambda_R^{\rm corr}$ are not an artifact of PSF correction, but rather evidence of a decrease of rotational support in low-redshift ETGs. Note that a less restrictive cut would not alter the results, whereas a more stringent cut would yield an even smaller MAGNUS subsample, making this test difficult to perform.\\

Since the samples are also matched in velocity dispersion, we repeated the above comparisons by binning galaxies with respect to $\sigma_e$. We sorted galaxies into bins of width $0.1$ dex in $\rm{log} \sigma_e$ and the same edge-bin treatment described above. Figure~\ref{fig:sigma_e_binned} shows the resulting medians of $\lambda_R^{\rm corr}$ for both samples, including the full sample (left) and the sample for which $\lambda_R^{\rm obs}/\lambda_R^{\rm corr} > 0.6$ (right). In both cases, the MAGNUS galaxies again display systematically higher $\lambda_R^{\rm corr}$ than their MaNGA counterparts. For the full samples, the average difference between the medians is $\sim65\%$, consistent with the stellar-mass based binning shown in Fig.~\ref{fig:lr_mass}. A sharp drop in $\lambda_R^{\rm corr}$ is still apparent around $\sigma_e \sim 200\,{\rm km\,s^{-1}}$. We find that in both MaNGA and MAGNUS samples, $70$--$80\%$ of galaxies with $\sigma_e > 200\,{\rm km\,s^{-1}}$ have $\log M_*/M_\odot \gtrsim 11.3$, and vice versa, indicating that the sharper decline seen at the high-mass end in Fig.~\ref{fig:lr_mass} and the corresponding feature near $\sigma_e \sim 200\,{\rm km\,s^{-1}}$ in Fig.~\ref{fig:sigma_e_binned} largely reflect the same subpopulation of galaxies. The findings for the full samples remains robust when restricting to the subsample least affected by the PSF correction.

Taken together, these results indicate that ETGs at intermediate redshifts possess higher rotational support than their low-redshift counterparts. This trend is consistent with an evolutionary pathway in which ETGs gradually lose angular momentum as they evolve from rotation-dominated systems into more pressure-supported spheroids toward the present day. The galaxy properties, for both MAGNUS and MaNGA samples, used in reaching this conclusion were measured using the same techniques and assumptions. Thus, potential systematic effects from the analysis were minimized, or at least affected both samples in the same way, further strengthening the robustness of our results. 

\begin{figure*}
    \centering
    {\includegraphics[width=\textwidth]{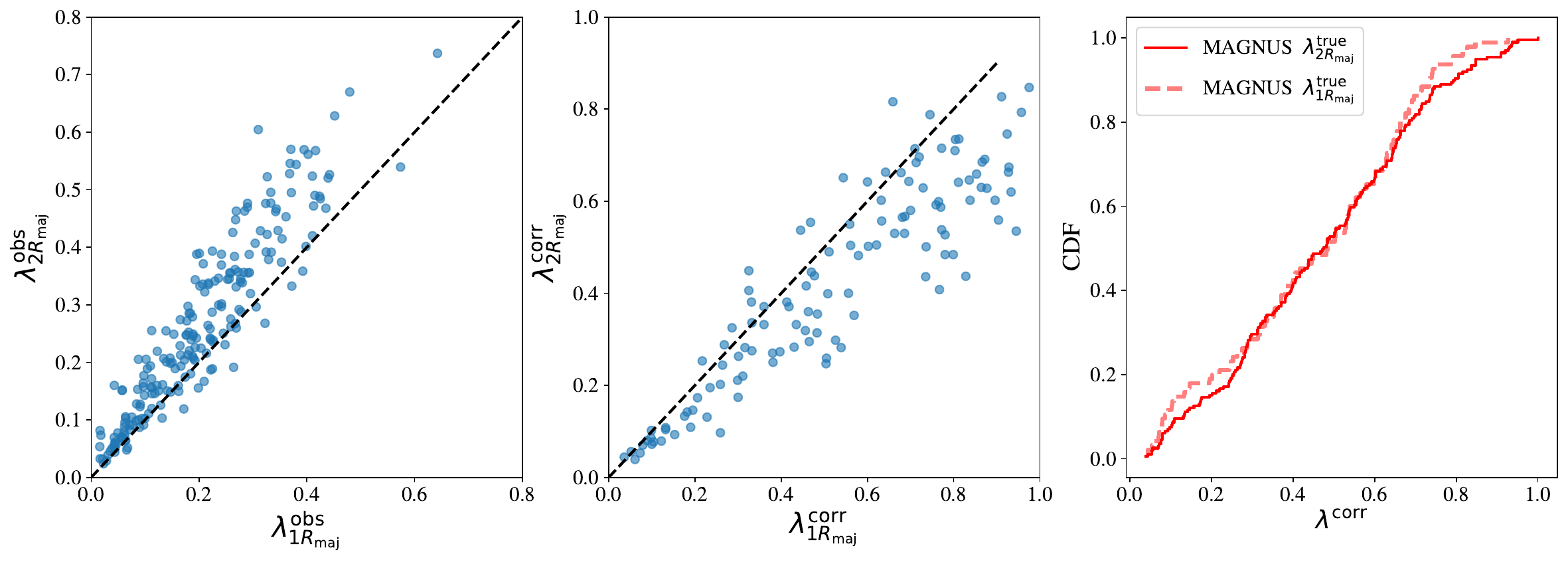}}
    \caption{Effect of aperture size on MAGNUS galaxies. Left: Comparison between the observed \lmr\ measured in elliptical apertures with a semi-major axis of 2 $R_e^{\rm maj}$ and 1 \remaj. The black dashed line is the one-to-one line. Middle: Comparison between PSF corrected \lmr\ from the two apertures. Here, we compare the PSF-corrected \lmr\ of 120 MAGNUS galaxies for which the correction could apply to observed values from both apertures. However, aperture size does not affect the galaxies with $\lmr\ < 0.15$, which are predominantly rounder and kinematically classified as slow rotators. Right: CDFs of the corrected \lmr\ distributions from both apertures.}
    \label{fig:comp_aper_lmr}
\end{figure*}

\section{Discussion}\label{sec:discussion}
One possible source of bias is aperture size, which affects the correction through the `resolution parameter', $r$. The observed \lmr\ values for MaNGA galaxies were measured using an elliptical aperture with a semi-major axis of 1 \remaj. As observed \lmr\ values usually increase with a bigger aperture, it may appear that the evidence of increased rotational support in MAGNUS galaxies is due to our adopted 2 $R_e^{\rm maj}$ elliptical aperture. While this statement is mostly correct for observed values, it is not true for PSF-corrected values, after correction.
In the leftmost plot of Fig.~\ref{fig:comp_aper_lmr}, we show the observed \lmr\ of MAGNUS galaxies measured in 1\remaj\ and 2\remaj\ apertures, and as expected, in most cases, observed \lmr\ from larger apertures is higher. However, the trend is opposite for corrected \lmr. In the middle plot of Fig.~\ref{fig:comp_aper_lmr}, we compare the corrected \lmr\ of 120 MAGNUS galaxies for which the correction was applicable for observed values from both apertures. For the rest of the galaxies, the correction could not be applied to either or both apertures because either $r > 1$ or $\lambda_{R}^{\rm corr} > 1$. We found that for most of these galaxies, $\lambda_{1 R_e^{\mathrm {maj}}}^{\mathrm{corr}} > \lambda_{2  R_e^{\mathrm {maj}}}^{\mathrm{corr}}$. Obviously, this is true for other galaxies for which the correction could not be applied. This means that, by using a smaller aperture, we would have found more differences between MAGNUS and MaNGA. In contrast, by adopting a larger aperture, we effectively minimized the effect of PSF correction. In the rightmost plot of Fig.~\ref{fig:comp_aper_lmr}, we show the CDFs of the corrected \lmr\ distributions from both apertures, which show that the distributions are especially different at both lower and upper ends.\\

The S{\'e}rsic index, $n$, also impacts the PSF correction. In general, a higher $n$ would compensate more than a lower $n$ at a fixed  $r$, and at poor resolution (large  $r$ ), the correction becomes even higher. That means if the S{\'e}rsic index of MAGNUS galaxies were systematically higher, that would also make the differences between MAGNUS and MaNGA higher at the adopted aperture size. The median S{\'e}rsic index of MAGNUS galaxies is 2.5, typical for ETGs at intermediate redshift, and we do not expect a systemic bias towards the lower side. On the other hand, if the S{\'e}rsic index of MaNGA galaxies were lower, thus less correction, it would also make the differences between MAGNUS and MaNGA higher. Only a systematically lower S{\'e}rsic index for MAGNUS and a higher one for MaNGA would reduce the observed differences between the samples. \\

The MAGNUS sample is statistically larger than previous samples of galaxies at similar redshifts with comparable data quality, yet it remains modest relative to the local MaNGA reference sample and is biased toward large, bright, and massive ETGs. Because the MaNGA set used here is mass matched to MAGNUS, fainter, smaller, and less massive systems that could follow different kinematic tracks are absent from both datasets. Substantial PSF corrections are unavoidable; although we adopt established prescriptions, residual systematics cannot be entirely excluded.\\

Progenitor bias \citep{vanDokkum_1996} also remains a consideration, since a mass-matched local sample need not be the exact descendant population of the high-$z$ ETGs. We do not correct for this effect explicitly, but we expect its impact to be small. If we ignore morphological properties (e.g., size) and stellar population properties (e.g., age and star-formation rate) and map only by stellar mass, such that a high-$z$ ETG evolves into a low-$z$ ETG of equal or higher mass, our conclusion is unchanged. The median \lmr\ trends as a function of mass are nearly parallel for the two surveys (see Fig.~\ref{fig:lr_mass}); taking any MAGNUS mass bin and the next higher MaNGA bin, the MaNGA bin exhibits an even lower \lmr\ than the corresponding MAGNUS bin. We also found that the same trend holds when comparing the two samples at fixed $\sigma_e$. For an evolving galaxy, the stellar mass is expected to be altered by mergers while its central velocity dispersion is considered a comparatively more stable property \citep[][and references therein]{LEGAC_Bezanson_2018}. Thus a comparison at fixed $\sigma_e$ is therefore less sensitive to assumptions about how galaxies move between mass bins across cosmic time, providing a further check against progenitor bias. Nonetheless, this limitation could still affect the amplitude of the inferred evolution and should be revisited with deeper, more complete samples. \\

Our findings agree with theoretical expectations from hydro-dynamical simulations \citep[e.g.][]{Lagos_2018a, Schulze_2018} and similar observational studies \citep[e.g.][]{LEGAC_Bezanson_2018, Derkenne_2024, DEugenio_2023}. \citet{LEGAC_Bezanson_2018} found statistically significant evidence for increased rotational support at $z\sim0.8$ relative to the local Universe, using slit spectroscopy. Our results, obtained independently with IFU data, support their conclusion. \citet{Derkenne_2024}, by contrast, found no significant evolution over the baseline probed by their study (0$<z\lesssim 0.3$). Given the short lookback time involved ($\sim$3--4 Gyr), this is consistent with only modest evolution having occurred by that epoch. \\

The physical drivers of the observed decline in rotational support point most directly to the
cumulative impact of mergers where both major and minor events reduce the specific angular momentum of galaxies \citep{Naab_2014,penoyre2017, Lagos_2018a}.
Interestingly, we find that the decline in $\lambda_R$ with redshift is not uniform across stellar mass. While ETGs in all of the used mass range show similar evolutionary trends, galaxies with $\log M_\ast > 11.3$ exhibit a noticeably steeper decrease in rotational support. A similar trend was also observed by \citet{Pulsoni_2023}. This is consistent with theoretical models predicting that angular momentum loss is most efficient in the most massive systems, which experience higher merger rates and more frequent dissipationless accretion through dry mergers \citep[e.g.][]{Rodriguez-Gomez_2017, Pulsoni_2021}. A comparably steep decline is also found among the highest-dispersion systems ($\sigma_e \gtrsim 200\,{\rm km\,s^{-1}}$) when binned by $\sigma_e$ instead of stellar mass. We further confirm that these are largely the same galaxies in both samples, where around $70$--$80\%$ of galaxies with $\sigma_e \gtrsim 200\,{\rm km\,s^{-1}}$ also satisfy $\log M_*/M_\odot \gtrsim 11.3$, and vice versa. The roughly parallel shift between MAGNUS and MaNGA samples at lower masses, and similarly at lower velocity dispersions, might suggest a more universal evolutionary mechanism, whereas the sharper decline in the high-mass, high-dispersion regime highlights
the importance of mass-dependent processes. \\

The agreement between our observations and theoretical predictions strengthens the case that ETGs at intermediate redshift are intrinsically more rotationally supported than their low-redshift counterparts. Moreover, the lack of significant changes in this trend when restricting to subsamples with minimal PSF corrections indicates that our findings are not driven by observational biases, but instead reflect a genuine physical difference. 

\section{Conclusion}\label{sec:conclusion}
In this work, we investigated the degree of rotational support in ETGs at different cosmic epochs, comparing high-redshift ETGs with their local counterparts. For our local reference, we used a sample of ETGs from the MaNGA survey, focusing on galaxies with redshifts below 0.05 and stellar masses of $\log M_{\ast} (M_{\odot}) \geq 10.6$. We compared intrinsic $\lambda_R$ values of this local sample with the ones of the MAGNUS sample of higher-redshift galaxies observed with MUSE ($0.25 < z < 0.75$), as $\lambda_R$ provides a direct measure of the balance between ordered rotation and random stellar motions in galaxies, and is therefore a key diagnostic of how rotational support evolves across cosmic time. We corrected the observed \lmr\ of MAGNUS galaxies for PSF effects following the same method applied to MaNGA ETGs. Although we used a bigger aperture to measure \lmr\ for MAGNUS galaxies to minimize resolution-related seeing correction than MaNGA, we argued that this does not affect our conclusions. Our analysis reveals that ETGs in the MAGNUS sample exhibit systematically higher $\lambda_R$ values compared to their local MaNGA counterparts. Here are our key findings : 

\begin{enumerate}
    \item The median value of $\lambda_R^{\rm corr}$ in the MAGNUS sample of 200 ETGs is $0.48 \pm 0.05$ while for the MaNGA sample it is $0.34 \pm 0.03$ and they originate from statistically distinguished populations.

    \item We also observed this declining trend in $z-\lambda_{R}$ relation. While the MAGNUS-only fit suggests a relatively shallow decline, $\mathrm{d} \lambda_R^{\rm corr}/ {\mathrm{d} z} = 0.10 \pm 0.12$, combining with MaNGA galaxies yields a steeper positive slope, $\mathrm{d} \lambda_R^{\rm corr}/ {\mathrm{d} z} = 0.25 \pm 0.04$, consistent with a net decrease in rotational support toward $z \sim 0$. The combined dataset spans a wider redshift baseline, thus more strongly highlighting the evolution across cosmic time. 


    \item The systematically higher rotational support of MAGNUS galaxies persists when the two samples are compared in bins of stellar mass and of velocity dispersion, with MAGNUS galaxies showing medians on average $\sim75\%$ higher than MaNGA across the stellar-mass range, and $\sim65\%$ higher across the velocity-dispersion range. In both cases, the difference is most pronounced among the most massive ($\log M_*/M_\odot > 11.3$) and highest-dispersion ($\sigma_e \gtrsim 200\,{\rm km\,s^{-1}}$) systems.

    \item This trend remains unchanged when the comparison is restricted to the subsamples of galaxies with $\lambda_R^{\rm obs}/\lambda_R^{\rm corr} > 0.6$, i.e. those least affected by the PSF correction, indicating that the result is not driven by the correction itself.

    \item The CDFs of $\lambda_R^{\rm corr}$ for the MAGNUS and MaNGA samples are statistically distinct, both for the full samples and for the subsamples minimally affected by the PSF correction ($p=0.001$ for the full samples). Within each survey, the full-sample and PSF-corrected-subsample distributions are themselves nearly indistinguishable ($p=0.25$ for MAGNUS), indicating that the correction does not disproportionately affect any particular subset of galaxies in either sample.
\end{enumerate}

These results confirm a decrease in rotational support over the last $\sim$ 7 Gyr in massive ETGs, consistent with the theoretical picture of spin-down in quiescent galaxies toward low redshift. The galaxy properties for both MAGNUS and MaNGA samples were measured using the same techniques and assumptions, minimizing the risk of systematic biases.  This analysis employed high-quality IFU data to derive \lmr\ and used a larger sample of both local and high-redshift galaxies compared to previous studies. This study enhances our understanding of the dynamical evolution of ETGs from high to low redshift, highlighting the importance of detailed kinematic studies in unraveling the complexities of galaxy evolution. Future IFU surveys extending to $z > 1$ will be crucial to constrain when this transition from rotation- to dispersion-dominated kinematics begins and how it depends on galaxy mass and velocity dispersion.

\begin{acknowledgments}
P.M. and T.T. acknowledge support from the National Science Foundation through grant NSF-AST-2407277. CDF acknowledges support for this work from the National Science Foundation under Grant No. AST-2407278.
\end{acknowledgments}

\textbf{Data availability}: The derived quantities for the MAGNUS sample used in this work, such as the S\'ersic index, corrected \lmr, and adopted PSF FWHM, are available in machine-readable format in the online journal. \\

%

\vspace{5mm}


\software{{\tt Python~3.12:} \citep{VanRossum2009}, 
{\tt Matplotlib~3.6:} \citep{Hunter2007}, 
{\tt NumPy~1.22:} \citep{Harris2020}, 
{\tt AstroPy~5.1} \citep{AstropyCollaboration2022}, 
{\tt pPXF~8.2} \citep{LEGAC_ppxf_Cappellari_2023}, 
{\tt MgeFit~5.0} \citep{MGE_Cappellari_2002}.
} 
 


\appendix

\section{Validation of pipeline provided PSF values}
\label{sec:appendix_psf}

To assess the reliability of the pipeline provided PSF FWHM values adopted in this work, we directly measured the PSF from field stars in a subset of the MUSE-DEEP datacubes and compared the results with the corresponding pipeline-provided values. We identified 18 relatively isolated, unsaturated stars across 5 datacubes. For each star, we used the \texttt{MoffatModel2.from\_starfit} function of the \textsc{mpdaf} package \citep{MPDAF_Bacon_2016} to characterize the PSF as follows: a spatial cutout of the datacube centered on the star was extracted, and this cutout was divided into a set of narrow wavelength bins spanning the MUSE wavelength range. A 2D circular Moffat profile was fitted independently to the collapsed image of each bin, yielding a FWHM and $\beta$ measurement per bin. A low-order polynomial was then fitted to these per-bin FWHM values as a function of wavelength, providing a smooth model FWHM($\lambda$) for each star. An analogous polynomial was also fitted to the per-bin $\beta$ measurements. This approach follows the empirical characterization of the MUSE PSF established during instrument commissioning, where the spatial point-spread function was found to be well described by a circular Moffat profile with a constant $\beta$ and a FWHM varying linearly with wavelength \citep{Bacon_2017}. For each star, we evaluated the resulting FWHM($\lambda$) polynomial over the observed-frame wavelength range 5000--8000~\AA, corresponding to the redshifted rest-frame window of 3780--4600~\AA\ used for stellar kinematic extraction across the MAGNUS sample redshift range $0.25 < z < 0.75$, and computed the mean of the model over this range. The uncertainty on this mean is taken as the RMS scatter of the individual per-bin Moffat measurements around the polynomial model. \\

\begin{figure}
    \centering
    {\includegraphics[width=0.95\columnwidth]{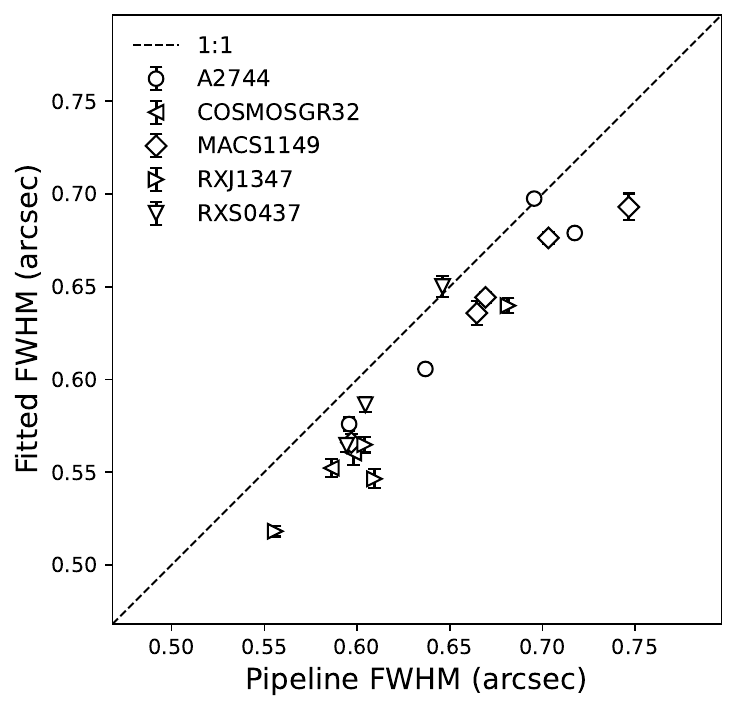}}
    \caption{Comparison between the pipeline provided and the directly-fitted PSF FWHM values for 18 field stars across 5 MUSE-DEEP datacubes. Each marker represents a particular cluster/field covered by the corresponding data cube. The black dashed line indicates the one-to-one relation. The pipeline values are systematically higher than the directly-fitted ones, with a mean offset of $5.1\%$ (median $5.4\%$, standard deviation $2.8\%$).}
    \label{fig:psf_validation}
\end{figure}

Table~\ref{tab:psf_validation} and Fig.~\ref{fig:psf_validation} compare these directly fitted FWHM values to the corresponding pipeline values, where the latter were computed as the average of the $x$- and $y$-directional PSF FWHM estimates at the star's position, provided by the corresponding datacube's header file, following the same procedure described in Sect.~\ref{sec:psf correction} for the galaxy sample. The two sets of measurements are in reasonable agreement, with the pipeline provided estimates being on average $5.1\%$ higher than the directly fitted ones (median $5.4\%$, standard deviation $2.8\%$). We additionally note that, within individual datacubes, stars at different positions show measurably different FWHM values, confirming that the effective seeing varies appreciably across the FOV of the combined MUSE-DEEP mosaics, consistent with our adoption of position-specific rather than global FWHM estimates.

To quantify the impact of this PSF FWHM offset on our results, we recomputed $\lambda_R^{\rm corr}$ for the full MAGNUS sample after reducing each galaxy's adopted $\sigma_{\rm PSF}$ by a uniform $5\%$, following the correction recipe of Sect.~\ref{sec:psf correction}. The revised median $\lambda_R^{\rm corr}$ is $0.46\pm0.03$, consistent within the uncertainty of the original measurement of $0.48\pm0.05$. To further account for the star-to-star variation in the measured PSF FWHM offset, we repeated the exercise by assigning each galaxy an independently drawn fractional $\sigma_{\rm PSF}$ correction sampled from a normal distribution with mean $5\%$ and standard deviation $3\%$, matching the distribution of offsets measured from the 18 field stars. Averaging over 1000 independent realizations of this per-galaxy sampling, the mean revised median is $0.46$, with negligible variation between realizations. In both tests, the shift in the sample median is well within the quoted statistical uncertainty and thus would not affect the conclusions of this paper.

\begin{table*}
\centering
\caption{Comparison of directly-fitted and pipeline provided PSF FWHM values for field stars in a subset of MUSE-DEEP datacubes used in this work.}
\label{tab:psf_validation}
\begin{tabular}{ccccccc}
\hline\hline
Cluster/Field & ESO Archieve ID & RA & Dec & Fitted FWHM & Pipeline FWHM & Difference \\
      &    & (J2000) & (J2000) & (arcsec) & (arcsec) & (\%)\\
\hline
A2744 & ADP.2017-03-24T12:14:09.100 &  00:14:19.48  &  -30:24:51.66  &  0.7  & 0.7 &  -0.29 \\
      &                             &  00:14:16.70  &  -30:24:18.16  &  0.68     & 0.72 &  5.66 \\
 &  &  00:14:19.37  &  -30:23:08.66  &  0.61     & 0.64 &  5.15 \\
 &  &  00:14:21.80  &  -30:24:09.16  &  0.58     & 0.6 &  3.44 \\
COSMOSGR32 & ADP.2019-01-02T10:06:51.914 &  09:59:41.13  &  02:30:23.9  &  0.56     & 0.6 &  6.75 \\
 &  &  09:59:41.57  &  02:30:21.6  &  0.55     & 0.59 &  6.15 \\
MACS2129  & ADP.2017-03-27T15:26:07.717 &  21:29:28.35  &  -07:41:46.2  &  0.68     & 0.7 &  3.98 \\
 &  &  21:29:28.72  &  -07:41:41.2  &  0.64     & 0.67 &  3.88 \\
 &  &  21:29:29.18  &  -07:41:13.8  &  0.69     & 0.75 &  7.73 \\
 &  &  21:29:24.69  &  -07:41:34.5  &  0.57     & 0.6 &  5.47 \\
 &  &  21:29:27.76  &  -07:41:19.0  &  0.64     & 0.66 &  4.53 \\
RXJ1347  & ADP.2019-01-02T09:51:36.307 &  13:47:28.68  &  -11:45:37.7  &  0.52     & 0.56 &  7.28 \\
 &  &  13:47:28.02  &  -11:44:33.9  &  0.56     & 0.6 &  6.96 \\
 &  &  13:47:27.97  &  -11:45:44.5  &  0.55     & 0.61 &  11.53 \\
 &  &  13:47:34.43  &  -11:44:30.9  &  0.64     & 0.68 &  6.5 \\
RXS0437  & ADP.2021-04-29T08:02:29.347 &  04:37:12.02  &  00:43:29.6  &  0.65     & 0.65 &  -0.61 \\
 &  &  04:37:06.83  &  00:43:49.0  &  0.56     & 0.59 &  5.29 \\
 &  &  04:37:08.31  &  00:43:59.1  &  0.59     & 0.6 &  3.08 \\
\hline
\end{tabular}
\tablecomments{Fitted FWHM values are the mean of the polynomial Moffat model (Sect.~\ref{sec:appendix_psf}) evaluated over the observed-frame range 5000--8000~\AA, with uncertainties given by the RMS scatter of the per-slice measurements around the polynomial. Pipeline provided FWHM values are taken from the header file of the corresponding data cube at the specific field position of the star. The percentage difference is computed as (pipeline $-$fitted)/fitted $\times$ 100.}
\end{table*}

\bibliographystyle{aasjournalv7}

\bibliography{pritom_magnus_ii}

@ARTICLE{Cappellari_2017_ppxf,
       author = {{Cappellari}, M.},
        title = "{Improving the full spectrum fitting method: accurate convolution with Gauss-Hermite functions}",
      journal = {MNRAS},
         year = {2017},
        month = apr,
       volume = {466},
       Ignore_number = {1},
        pages = {798-811},
          doi = {10.1093/mnras/stw3020},
archivePrefix = {arXiv},
       eprint = {1607.08538},
 primaryClass = {astro-ph.GA},
       adsurl = {https://ui.adsabs.harvard.edu/abs/2017MNRAS.466..798C}
}

@ARTICLE{Moffat_1969,
       author = {{Moffat}, A.~F.~J.},
        title = "{A Theoretical Investigation of Focal Stellar Images in the Photographic Emulsion and Application to Photographic Photometry}",
      journal = {\aap},
         year = 1969,
        month = dec,
       volume = {3},
        pages = {455},
       adsurl = {https://ui.adsabs.harvard.edu/abs/1969A&A.....3..455M}
}

@ARTICLE{Indo_US_lib,
       author = {{Valdes}, Francisco and {Gupta}, Ranjan and {Rose}, James A. and {Singh}, Harinder P. and {Bell}, David J.},
        title = "{The Indo-US Library of Coud{\'e} Feed Stellar Spectra}",
      journal = {\apjs},
         year = 2004,
        month = jun,
       volume = {152},
       number = {2},
        pages = {251-259},
          doi = {10.1086/386343},
archivePrefix = {arXiv},
       eprint = {astro-ph/0402435},
 primaryClass = {astro-ph},
       adsurl = {https://ui.adsabs.harvard.edu/abs/2004ApJS..152..251V}
}

@ARTICLE{MILES_library_2011,
       author = {{Beifiori}, A. and {Maraston}, C. and {Thomas}, D. and {Johansson}, J.},
        title = "{On the spectral resolution of the MILES stellar library}",
      journal = {\aap},
         year = 2011,
        month = jul,
       volume = {531},
          eid = {A109},
        pages = {A109},
          doi = {10.1051/0004-6361/201016323},
archivePrefix = {arXiv},
       eprint = {1012.3428},
 primaryClass = {astro-ph.CO},
       adsurl = {https://ui.adsabs.harvard.edu/abs/2011A&A...531A.109B}
}

@Article{Hunter2007,
  author    = {Hunter, J. D.},
  journal   = {Computing in Science and Engineering},
  title     = {Matplotlib: A 2D Graphics Environment},
  year      = {2007},
  month     = may,
  number    = {3},
  pages     = {90-95},
  volume    = {9},
  doi       = {10.1109/MCSE.2007.55},
  publisher = {IEEE COMPUTER SOC},
  url       = {http://adsabs.harvard.edu/abs/2007CSE.....9...90H},
}

@ARTICLE{ATLAS_III_Emselem_2011,
       author = {{Emsellem}, Eric and {Cappellari}, Michele and {Krajnovi{\'c}}, Davor and {Alatalo}, Katherine and {Blitz}, Leo and {Bois}, Maxime and {Bournaud}, Fr{\'e}d{\'e}ric and {Bureau}, Martin and {Davies}, Roger L. and {Davis}, Timothy A. and {de Zeeuw}, P.~T. and {Khochfar}, Sadegh and {Kuntschner}, Harald and {Lablanche}, Pierre-Yves and {McDermid}, Richard M. and {Morganti}, Raffaella and {Naab}, Thorsten and {Oosterloo}, Tom and {Sarzi}, Marc and {Scott}, Nicholas and {Serra}, Paolo and {van de Ven}, Glenn and {Weijmans}, Anne-Marie and {Young}, Lisa M.},
        title = "{The ATLAS$^{3D}$ project - III. A census of the stellar angular momentum within the effective radius of early-type galaxies: unveiling the distribution of fast and slow rotators}",
      journal = {\mnras},
         year = 2011,
        month = jun,
       volume = {414},
       number = {2},
        pages = {888-912},
          doi = {10.1111/j.1365-2966.2011.18496.x},
archivePrefix = {arXiv},
       eprint = {1102.4444},
 primaryClass = {astro-ph.CO},
       adsurl = {https://ui.adsabs.harvard.edu/abs/2011MNRAS.414..888E}
}

@ARTICLE{ATLAS_XV_Michele_2013,
       author = {{Cappellari}, Michele and {Scott}, Nicholas and {Alatalo}, Katherine and {Blitz}, Leo and {Bois}, Maxime and {Bournaud}, Fr{\'e}d{\'e}ric and {Bureau}, M. and {Crocker}, Alison F. and {Davies}, Roger L. and {Davis}, Timothy A. and {de Zeeuw}, P.~T. and {Duc}, Pierre-Alain and {Emsellem}, Eric and {Khochfar}, Sadegh and {Krajnovi{\'c}}, Davor and {Kuntschner}, Harald and {McDermid}, Richard M. and {Morganti}, Raffaella and {Naab}, Thorsten and {Oosterloo}, Tom and {Sarzi}, Marc and {Serra}, Paolo and {Weijmans}, Anne-Marie and {Young}, Lisa M.},
        title = "{The ATLAS$^{3D}$ project - XV. Benchmark for early-type galaxies scaling relations from 260 dynamical models: mass-to-light ratio, dark matter, Fundamental Plane and Mass Plane}",
      journal = {\mnras},
         year = 2013,
        month = jul,
       volume = {432},
       number = {3},
        pages = {1709-1741},
          doi = {10.1093/mnras/stt562},
archivePrefix = {arXiv},
       eprint = {1208.3522},
 primaryClass = {astro-ph.CO},
       adsurl = {https://ui.adsabs.harvard.edu/abs/2013MNRAS.432.1709C}
}

@ARTICLE{Bacon_2017,
   author = {{Bacon}, R. and {Conseil}, S. and {Mary}, D. and {Brinchmann}, J. and others},
    title = "{The MUSE Hubble Ultra Deep Field Survey. I. Survey description, data reduction, and source detection}",
  journal = {\aap},
     year = 2017,
    volume = 608,
     pages = {A1},
      doi = {10.1051/0004-6361/201731128},
   adsurl = {https://ui.adsabs.harvard.edu/abs/2017A&A...608A...1B}
}

@ARTICLE{Belli_2017,
       author = {{Belli}, Sirio and {Newman}, Andrew B. and {Ellis}, Richard S.},
        title = "{MOSFIRE Spectroscopy of Quiescent Galaxies at 1.5 < z < 2.5. I. Evolution of Structural and Dynamical Properties}",
      journal = {\apj},
         year = 2017,
        month = jan,
       volume = {834},
       number = {1},
          eid = {18},
        pages = {18},
          doi = {10.3847/1538-4357/834/1/18},
archivePrefix = {arXiv},
       eprint = {1608.00608},
 primaryClass = {astro-ph.GA},
       adsurl = {https://ui.adsabs.harvard.edu/abs/2017ApJ...834...18B}
}

@ARTICLE{CALIFA_survey_2012,
       author = {{S{\'a}nchez}, S.~F. and {Kennicutt}, R.~C. and {Gil de Paz}, A. and {van de Ven}, G. and {V{\'\i}lchez}, J.~M. and {Wisotzki}, L. and {Walcher}, C.~J. and {Mast}, D. and {Aguerri}, J.~A.~L. and {Albiol-P{\'e}rez}, S. and {Alonso-Herrero}, A. and {Alves}, J. and {Bakos}, J. and {Bart{\'a}kov{\'a}}, T. and {Bland-Hawthorn}, J. and {Boselli}, A. and {Bomans}, D.~J. and {Castillo-Morales}, A. and {Cortijo-Ferrero}, C. and {de Lorenzo-C{\'a}ceres}, A. and {Del Olmo}, A. and {Dettmar}, R. -J. and {D{\'\i}az}, A. and {Ellis}, S. and {Falc{\'o}n-Barroso}, J. and {Flores}, H. and {Gallazzi}, A. and {Garc{\'\i}a-Lorenzo}, B. and {Gonz{\'a}lez Delgado}, R. and {Gruel}, N. and {Haines}, T. and {Hao}, C. and {Husemann}, B. and {Igl{\'e}sias-P{\'a}ramo}, J. and {Jahnke}, K. and {Johnson}, B. and {Jungwiert}, B. and {Kalinova}, V. and {Kehrig}, C. and {Kupko}, D. and {L{\'o}pez-S{\'a}nchez}, {\'A}. R. and {Lyubenova}, M. and {Marino}, R.~A. and {M{\'a}rmol-Queralt{\'o}}, E. and {M{\'a}rquez}, I. and {Masegosa}, J. and {Meidt}, S. and {Mendez-Abreu}, J. and {Monreal-Ibero}, A. and {Montijo}, C. and {Mour{\~a}o}, A.~M. and {Palacios-Navarro}, G. and {Papaderos}, P. and {Pasquali}, A. and {Peletier}, R. and {P{\'e}rez}, E. and {P{\'e}rez}, I. and {Quirrenbach}, A. and {Rela{\~n}o}, M. and {Rosales-Ortega}, F.~F. and {Roth}, M.~M. and {Ruiz-Lara}, T. and {S{\'a}nchez-Bl{\'a}zquez}, P. and {Sengupta}, C. and {Singh}, R. and {Stanishev}, V. and {Trager}, S.~C. and {Vazdekis}, A. and {Viironen}, K. and {Wild}, V. and {Zibetti}, S. and {Ziegler}, B.},
        title = "{CALIFA, the Calar Alto Legacy Integral Field Area survey. I. Survey presentation}",
      journal = {\aap},
         year = 2012,
        month = feb,
       volume = {538},
          eid = {A8},
        pages = {A8},
          doi = {10.1051/0004-6361/201117353},
archivePrefix = {arXiv},
       eprint = {1111.0962},
 primaryClass = {astro-ph.CO},
       adsurl = {https://ui.adsabs.harvard.edu/abs/2012A&A...538A...8S}
}

@ARTICLE{Cappellari_review_2016,
       author = {{Cappellari}, Michele},
        title = "{Structure and Kinematics of Early-Type Galaxies from Integral Field Spectroscopy}",
      journal = {\araa},
         year = 2016,
        month = sep,
       volume = {54},
        pages = {597-665},
          doi = {10.1146/annurev-astro-082214-122432},
archivePrefix = {arXiv},
       eprint = {1602.04267},
 primaryClass = {astro-ph.GA},
       adsurl = {https://ui.adsabs.harvard.edu/abs/2016ARA&A..54..597C}
}

@article{Crain_2015_EAGLE,
  author = {Crain, R. A. and Schaye, J. and Bower, R. G. and Furlong, M. and Schaller, M. and Theuns, T. and Dalla Vecchia, C. and Frenk, C. S. and McCarthy, I. G. and Helly, J. C. and Jenkins, A. and Rosas-Guevara, Y. M. and White, S. D. M. and Baes, M. and Booth, C. M. and Camps, P. and Navarro, J. F. and Qu, Y. and Rahmati, A. and Sawala, T. and Thomas, P. A. and Trayford, J. W.},
  title = {The EAGLE simulations of galaxy formation: calibration of subgrid physics and model variations},
  journal = {Monthly Notices of the Royal Astronomical Society},
  volume = {450},
  number = {2},
  pages = {1937--1961},
  year = {2015},
  doi = {10.1093/mnras/stv725}
}

@ARTICLE{Derkenne_2024,
       author = {{Derkenne}, Caro and {McDermid}, Richard M. and {D'Eugenio}, Francesco and {Foster}, Caroline and {Khalid}, Aman and {Harborne}, Katherine E. and {van de Sande}, Jesse and {Croom}, Scott M. and {Lagos}, Claudia D.~P. and {Bellstedt}, Sabine and {Mendel}, J. Trevor and {Mun}, Marcie and {Wisnioski}, Emily and {Bagge}, Ryan S. and {Battisti}, Andrew J. and {Bland-Hawthorn}, Joss and {Ferr{\'e}-Mateu}, Anna and {Peng}, Yingjie and {Santucci}, Giulia and {Sweet}, Sarah M. and {Thater}, Sabine and {Valenzuela}, Lucas M. and {Ziegler}, Bodo},
        title = "{The MAGPI Survey: massive slow rotator population in place by z   0.3}",
      journal = {\mnras},
         year = 2024,
        month = jul,
       volume = {531},
       number = {4},
        pages = {4602-4610},
          doi = {10.1093/mnras/stae1407},
archivePrefix = {arXiv},
       eprint = {2406.02185},
 primaryClass = {astro-ph.GA},
       adsurl = {https://ui.adsabs.harvard.edu/abs/2024MNRAS.531.4602D}
}

@ARTICLE{DEugenio_2023,
       author = {{D'Eugenio}, Francesco and {van der Wel}, Arjen and {Piotrowska}, Joanna M. and {Bezanson}, Rachel and {Taylor}, Edward N. and {van de Sande}, Jesse and {Baker}, William M. and {Bell}, Eric F. and {Bellstedt}, Sabine and {Bland-Hawthorn}, Joss and {Bluck}, Asa F.~L. and {Brough}, Sarah and {Bryant}, Julia J. and {Colless}, Matthew and {Cortese}, Luca and {Croom}, Scott M. and {Derkenne}, Caro and {van Dokkum}, Pieter and {Fisher}, Deanne and {Foster}, Caroline and {Gallazzi}, Anna and {de Graaff}, Anna and {Groves}, Brent and {van Houdt}, Josha and {Lagos}, Claudia del P. and {Looser}, Tobias J. and {Maiolino}, Roberto and {Maseda}, Michael and {Mendel}, J. Trevor and {Nersesian}, Angelos and {Pacifici}, Camilla and {Poci}, Adriano and {Remus}, Rhea-Silvia and {Sweet}, Sarah M. and {Thater}, Sabine and {Tran}, Kim-Vy and {{\"U}bler}, Hannah and {Valenzuela}, Lucas M. and {Wisnioski}, Emily and {Zibetti}, Stefano},
        title = "{Evolution in the orbital structure of quiescent galaxies from MAGPI, LEGA-C, and SAMI surveys: direct evidence for merger-driven growth over the last 7 Gyr}",
      journal = {\mnras},
         year = 2023,
        month = oct,
       volume = {525},
       number = {2},
        pages = {2789-2805},
          doi = {10.1093/mnras/stad800},
archivePrefix = {arXiv},
       eprint = {2303.05520},
 primaryClass = {astro-ph.GA},
       adsurl = {https://ui.adsabs.harvard.edu/abs/2023MNRAS.525.2789D}
}

@ARTICLE{Dynpop_I_Zhu_2023,
       author = {{Zhu}, Kai and {Lu}, Shengdong and {Cappellari}, Michele and {Li}, Ran and {Mao}, Shude and {Gao}, Liang},
        title = "{MaNGA DynPop - I. Quality-assessed stellar dynamical modelling from integral-field spectroscopy of 10K nearby galaxies: a catalogue of masses, mass-to-light ratios, density profiles, and dark matter}",
      journal = {\mnras},
         year = 2023,
        month = jul,
       volume = {522},
       number = {4},
        pages = {6326-6353},
          doi = {10.1093/mnras/stad1299},
archivePrefix = {arXiv},
       eprint = {2304.11711},
 primaryClass = {astro-ph.GA},
       adsurl = {https://ui.adsabs.harvard.edu/abs/2023MNRAS.522.6326Z}
}

@ARTICLE{DynPop_II_Lu_2023,
       author = {{Lu}, Shengdong and {Zhu}, Kai and {Cappellari}, Michele and {Li}, Ran and {Mao}, Shude and {Xu}, Dandan},
        title = "{MaNGA DynPop - II. Global stellar population, gradients, and star-formation histories from integral-field spectroscopy of 10K galaxies: link with galaxy rotation, shape, and total-density gradients}",
      journal = {\mnras},
         year = 2023,
        month = nov,
       volume = {526},
       number = {1},
        pages = {1022-1045},
          doi = {10.1093/mnras/stad2732},
archivePrefix = {arXiv},
       eprint = {2304.11712},
 primaryClass = {astro-ph.GA},
       adsurl = {https://ui.adsabs.harvard.edu/abs/2023MNRAS.526.1022L}
}

@ARTICLE{FSPS_Conroy_2009,
       author = {{Conroy}, Charlie and {Gunn}, James E. and {White}, Martin},
        title = "{The Propagation of Uncertainties in Stellar Population Synthesis Modeling. I. The Relevance of Uncertain Aspects of Stellar Evolution and the Initial Mass Function to the Derived Physical Properties of Galaxies}",
      journal = {\apj},
         year = 2009,
        month = jul,
       volume = {699},
       number = {1},
        pages = {486-506},
          doi = {10.1088/0004-637X/699/1/486},
archivePrefix = {arXiv},
       eprint = {0809.4261},
 primaryClass = {astro-ph},
       adsurl = {https://ui.adsabs.harvard.edu/abs/2009ApJ...699..486C}
}

@ARTICLE{FSPS_Conroy_2010,
       author = {{Conroy}, Charlie and {Gunn}, James E.},
        title = "{The Propagation of Uncertainties in Stellar Population Synthesis Modeling. III. Model Calibration, Comparison, and Evaluation}",
      journal = {\apj},
         year = 2010,
        month = apr,
       volume = {712},
       number = {2},
        pages = {833-857},
          doi = {10.1088/0004-637X/712/2/833},
archivePrefix = {arXiv},
       eprint = {0911.3151},
 primaryClass = {astro-ph.CO},
       adsurl = {https://ui.adsabs.harvard.edu/abs/2010ApJ...712..833C}
}

@ARTICLE{GaLight_Xuheng_2021,
       author = {{Ding}, Xuheng and {Birrer}, Simon and {Treu}, Tommaso and {Silverman}, John D.},
        title = "{Galaxy shapes of Light (GaLight): a 2D modeling of galaxy images}",
      journal = {arXiv e-prints},
         year = 2021,
        month = nov,
          eid = {arXiv:2111.08721},
        pages = {arXiv:2111.08721},
          doi = {10.48550/arXiv.2111.08721},
archivePrefix = {arXiv},
       eprint = {2111.08721},
 primaryClass = {astro-ph.GA},
       adsurl = {https://ui.adsabs.harvard.edu/abs/2021arXiv211108721D}
}

@ARTICLE{Graham_2018,
       author = {{Graham}, Mark T. and {Cappellari}, Michele and {Li}, Hongyu and {Mao}, Shude and {Bershady}, Matthew A. and {Bizyaev}, Dmitry and {Brinkmann}, Jonathan and {Brownstein}, Joel R. and {Bundy}, Kevin and {Drory}, Niv and {Law}, David R. and {Pan}, Kaike and {Thomas}, Daniel and {Wake}, David A. and {Weijmans}, Anne-Marie and {Westfall}, Kyle B. and {Yan}, Renbin},
        title = "{SDSS-IV MaNGA: stellar angular momentum of about 2300 galaxies: unveiling the bimodality of massive galaxy properties}",
      journal = {\mnras},
         year = 2018,
        month = jul,
       volume = {477},
       number = {4},
        pages = {4711-4737},
          doi = {10.1093/mnras/sty504},
archivePrefix = {arXiv},
       eprint = {1802.08213},
 primaryClass = {astro-ph.GA},
       adsurl = {https://ui.adsabs.harvard.edu/abs/2018MNRAS.477.4711G}
}

@article{Hopkins_2009,
  author       = {Hopkins, Philip F. and Bundy, Kevin and Murray, Norman and Quataert, Eliot and Lauer, Tod R. and Ma, Chung-Pei},
  title        = {Compact High-Redshift Galaxies Are the Cores of the Most Massive Present-Day Spheroids},
  journal      = {Monthly Notices of the Royal Astronomical Society},
  volume       = {398},
  number       = {2},
  pages        = {898--922},
  year         = {2009},
  doi          = {10.1111/j.1365-2966.2009.15191.x}
}

@ARTICLE{Jampy_Cappellari_2008,
       author = {{Cappellari}, Michele},
        title = "{Measuring the inclination and mass-to-light ratio of axisymmetric galaxies via anisotropic Jeans models of stellar kinematics}",
      journal = {\mnras},
         year = 2008,
        month = oct,
       volume = {390},
       number = {1},
        pages = {71-86},
          doi = {10.1111/j.1365-2966.2008.13754.x},
archivePrefix = {arXiv},
       eprint = {0806.0042},
 primaryClass = {astro-ph},
       adsurl = {https://ui.adsabs.harvard.edu/abs/2008MNRAS.390...71C}
}

@ARTICLE{K-correction_Hogg_2002,
       author = {{Hogg}, David W. and {Baldry}, Ivan K. and {Blanton}, Michael R. and {Eisenstein}, Daniel J.},
        title = "{The K correction}",
      journal = {arXiv e-prints},
         year = 2002,
        month = oct,
          eid = {astro-ph/0210394},
        pages = {astro-ph/0210394},
          doi = {10.48550/arXiv.astro-ph/0210394},
archivePrefix = {arXiv},
       eprint = {astro-ph/0210394},
 primaryClass = {astro-ph},
       adsurl = {https://ui.adsabs.harvard.edu/abs/2002astro.ph.10394H}
}

@ARTICLE{Kinematic_classification_Emsellem_2007,
       author = {{Emsellem}, Eric and {Cappellari}, Michele and {Krajnovi{\'c}}, Davor and {van de Ven}, Glenn and {Bacon}, R. and {Bureau}, M. and {Davies}, Roger L. and {de Zeeuw}, P.~T. and {Falc{\'o}n-Barroso}, Jes{\'u}s and {Kuntschner}, Harald and {McDermid}, Richard and {Peletier}, Reynier F. and {Sarzi}, Marc},
        title = "{The SAURON project - IX. A kinematic classification for early-type galaxies}",
      journal = {\mnras},
         year = 2007,
        month = aug,
       volume = {379},
       number = {2},
        pages = {401-417},
          doi = {10.1111/j.1365-2966.2007.11752.x},
archivePrefix = {arXiv},
       eprint = {astro-ph/0703531},
 primaryClass = {astro-ph},
       adsurl = {https://ui.adsabs.harvard.edu/abs/2007MNRAS.379..401E}
}

@ARTICLE{Knabel_Mozumdar_2025,
       author = {{Knabel}, Shawn and {Mozumdar}, Pritom and {Shajib}, Anowar J. and {Treu}, Tommaso and {Cappellari}, Michele and {Spiniello}, Chiara and {Birrer}, Simon},
        title = "{TDCOSMO XIX: Measuring stellar velocity dispersion with sub-percent accuracy for cosmography}",
      journal = {arXiv e-prints},
         year = 2025,
        month = feb,
          eid = {arXiv:2502.16034},
        pages = {arXiv:2502.16034},
          doi = {10.48550/arXiv.2502.16034},
archivePrefix = {arXiv},
       eprint = {2502.16034},
 primaryClass = {astro-ph.GA},
       adsurl = {https://ui.adsabs.harvard.edu/abs/2025arXiv250216034K}
}

@ARTICLE{Lagos_2018b,
       author = {{Lagos}, Claudia del P. and {Schaye}, Joop and {Bah{\'e}}, Yannick and {van de Sande}, Jesse and {Kay}, Scott T. and {Barnes}, David and {Davis}, Timothy A. and {Dalla Vecchia}, Claudio},
        title = "{The connection between mass, environment, and slow rotation in simulated galaxies}",
      journal = {\mnras},
         year = 2018,
        month = jun,
       volume = {476},
       number = {4},
        pages = {4327-4345},
          doi = {10.1093/mnras/sty489},
archivePrefix = {arXiv},
       eprint = {1712.01398},
 primaryClass = {astro-ph.GA},
       adsurl = {https://ui.adsabs.harvard.edu/abs/2018MNRAS.476.4327L}
}

@ARTICLE{Lagos_2018a,
       author = {{Lagos}, Claudia del P. and {Stevens}, Adam R.~H. and {Bower}, Richard G. and {Davis}, Timothy A. and {Contreras}, Sergio andSch {Padilla}, Nelson D. and {Obreschkow}, Danail and {Croton}, Darren and {Trayford}, James W. and {Welker}, Charlotte and {Theuns}, Tom},
        title = "{Quantifying the impact of mergers on the angular momentum of simulated galaxies}",
      journal = {\mnras},
         year = 2018,
        month = feb,
       volume = {473},
       number = {4},
        pages = {4956-4974},
          doi = {10.1093/mnras/stx2667},
archivePrefix = {arXiv},
       eprint = {1701.04407},
 primaryClass = {astro-ph.GA},
       adsurl = {https://ui.adsabs.harvard.edu/abs/2018MNRAS.473.4956L}
}

@ARTICLE{LEGAC_survey_Wel_2016,
       author = {{van der Wel}, A. and {Noeske}, K. and {Bezanson}, R. and {Pacifici}, C. and {Gallazzi}, A. and {Franx}, M. and {Mu{\~n}oz-Mateos}, J.~C. and {Bell}, E.~F. and {Brammer}, G. and {Charlot}, S. and {Chauk{\'e}}, P. and {Labb{\'e}}, I. and {Maseda}, M.~V. and {Muzzin}, A. and {Rix}, H. -W. and {Sobral}, D. and {van de Sande}, J. and {van Dokkum}, P.~G. and {Wild}, V. and {Wolf}, C.},
        title = "{The VLT LEGA-C Spectroscopic Survey: The Physics of Galaxies at a Lookback Time of 7 Gyr}",
      journal = {\apjs},
         year = 2016,
        month = apr,
       volume = {223},
       number = {2},
          eid = {29},
        pages = {29},
          doi = {10.3847/0067-0049/223/2/29},
archivePrefix = {arXiv},
       eprint = {1603.05479},
 primaryClass = {astro-ph.GA},
       adsurl = {https://ui.adsabs.harvard.edu/abs/2016ApJS..223...29V}
}

@ARTICLE{LEGAC_Bezanson_2018,
       author = {{Bezanson}, Rachel and {van der Wel}, Arjen and {Pacifici}, Camilla and {Noeske}, Kai and {Bari{\v{s}}i{\'c}}, Ivana and {Bell}, Eric F. and {Brammer}, Gabriel B. and {Calhau}, Joao and {Chauke}, Priscilla and {van Dokkum}, Pieter and {Franx}, Marijn and {Gallazzi}, Anna and {van Houdt}, Josha and {Labb{\'e}}, Ivo and {Maseda}, Michael V. and {Mu{\~n}os-Mateos}, Juan Carlos and {Muzzin}, Adam and {van de Sande}, Jesse and {Sobral}, David and {Straatman}, Caroline and {Wu}, Po-Feng},
        title = "{Spatially Resolved Stellar Kinematics from LEGA-C: Increased Rotational Support in z {\ensuremath{\sim}} 0.8 Quiescent Galaxies}",
      journal = {\apj},
         year = 2018,
        month = may,
       volume = {858},
       number = {1},
          eid = {60},
        pages = {60},
          doi = {10.3847/1538-4357/aabc55},
archivePrefix = {arXiv},
       eprint = {1804.02402},
 primaryClass = {astro-ph.GA},
       adsurl = {https://ui.adsabs.harvard.edu/abs/2018ApJ...858...60B}
}

@ARTICLE{LEGAC_ppxf_Cappellari_2023,
       author = {{Cappellari}, Michele},
        title = "{Full spectrum fitting with photometry in PPXF: stellar population versus dynamical masses, non-parametric star formation history and metallicity for 3200 LEGA-C galaxies at redshift z {\ensuremath{\approx}} 0.8}",
      journal = {\mnras},
         year = 2023,
        month = dec,
       volume = {526},
       number = {3},
        pages = {3273-3300},
          doi = {10.1093/mnras/stad2597},
archivePrefix = {arXiv},
       eprint = {2208.14974},
 primaryClass = {astro-ph.GA},
       adsurl = {https://ui.adsabs.harvard.edu/abs/2023MNRAS.526.3273C}
}

@ARTICLE{Lenstronomy_II_2021,
   author = {{Birrer}, Simon and {Shajib}, Anowar and {Gilman}, Daniel and {Galan}, Aymeric and {Aalbers}, Jelle and {Millon}, Martin and {Morgan}, Robert and {Pagano}, Giulia and {Park}, Ji and {Teodori}, Luca and {Tessore}, Nicolas and {Ueland}, Madison and {Van de Vyvere}, Lyne and {Wagner-Carena}, Sebastian and {Wempe}, Ewoud and {Yang}, Lilan and {Ding}, Xuheng and {Schmidt}, Thomas and {Sluse}, Dominique and {Zhang}, Ming and {Amara}, Adam},
    title = "{lenstronomy II: A gravitational lensing software ecosystem}",
  journal = {The Journal of Open Source Software},
     year = 2021,
    month = jun,
   volume = {6},
   number = {62},
      eid = {3283},
    pages = {3283},
      doi = {10.21105/joss.03283},
    archivePrefix = {arXiv},
   eprint = {2106.05976},
     primaryClass = {astro-ph.CO},
   adsurl = {https://ui.adsabs.harvard.edu/abs/2021JOSS....6.3283B}
    }

@ARTICLE{MAGPI_survey_design_Foster_2021,
       author = {{Foster}, C. and {Mendel}, J.~T. and {Lagos}, C.~D.~P. and {Wisnioski}, E. and {Yuan}, T. and {D'Eugenio}, F. and {Barone}, T.~M. and {Harborne}, K.~E. and {Vaughan}, S.~P. and {Schulze}, F. and {Remus}, R.-S. and {Gupta}, A. and {Collacchioni}, F. and {Khim}, D.~J. and {Taylor}, P. and {Bassett}, R. and {Croom}, S.~M. and {McDermid}, R.~M. and {Poci}, A. and {Battisti}, A.~J. and {Bland-Hawthorn}, J. and {Bellstedt}, S. and {Colless}, M. and {Davies}, L.~J.~M. and {Derkenne}, C. and {Driver}, S. and {Ferr{\'e}-Mateu}, A. and {Fisher}, D.~B. and {Gjergo}, E. and {Johnston}, E.~J. and {Khalid}, A. and {Kobayashi}, C. and {Oh}, S. and {Peng}, Y. and {Robotham}, A.~S.~G. and {Sharda}, P. and {Sweet}, S.~M. and {Taylor}, E.~N. and {Tran}, K.-V.~H. and {Trayford}, J.~W. and {van de Sande}, J. and {Yi}, S.~K. and {Zanisi}, L.},
        title = "{The MAGPI survey: Science goals, design, observing strategy, early results and theoretical framework}",
      journal = {\pasa},
         year = 2021,
        month = jul,
       volume = {38},
          eid = {e031},
        pages = {e031},
          doi = {10.1017/pasa.2021.25},
archivePrefix = {arXiv},
       eprint = {2011.13567},
 primaryClass = {astro-ph.GA},
       adsurl = {https://ui.adsabs.harvard.edu/abs/2021PASA...38...31F}
}

@ARTICLE{MaNGA_DAP_Westfall_2019,
       author = {{Westfall}, Kyle B. and {Cappellari}, Michele and {Bershady}, Matthew A. and {Bundy}, Kevin and {Belfiore}, Francesco and {Ji}, Xihan and {Law}, David R. and {Schaefer}, Adam and {Shetty}, Shravan and {Tremonti}, Christy A. and {Yan}, Renbin and {Andrews}, Brett H. and {Brownstein}, Joel R. and {Cherinka}, Brian and {Coccato}, Lodovico and {Drory}, Niv and {Maraston}, Claudia and {Parikh}, Taniya and {S{\'a}nchez-Gallego}, Jos{\'e} R. and {Thomas}, Daniel and {Weijmans}, Anne-Marie and {Barrera-Ballesteros}, Jorge and {Du}, Cheng and {Goddard}, Daniel and {Li}, Niu and {Masters}, Karen and {Ibarra Medel}, H{\'e}ctor Javier and {S{\'a}nchez}, Sebasti{\'a}n F. and {Yang}, Meng and {Zheng}, Zheng and {Zhou}, Shuang},
        title = "{The Data Analysis Pipeline for the SDSS-IV MaNGA IFU Galaxy Survey: Overview}",
      journal = {\aj},
         year = 2019,
        month = dec,
       volume = {158},
       number = {6},
          eid = {231},
        pages = {231},
          doi = {10.3847/1538-3881/ab44a2},
archivePrefix = {arXiv},
       eprint = {1901.00856},
 primaryClass = {astro-ph.GA},
       adsurl = {https://ui.adsabs.harvard.edu/abs/2019AJ....158..231W}
}

@ARTICLE{MaNGA_DRP_Law_2016,
       author = {{Law}, David R. and {Cherinka}, Brian and {Yan}, Renbin and {Andrews}, Brett H. and {Bershady}, Matthew A. and {Bizyaev}, Dmitry and {Blanc}, Guillermo A. and {Blanton}, Michael R. and {Bolton}, Adam S. and {Brownstein}, Joel R. and {Bundy}, Kevin and {Chen}, Yanmei and {Drory}, Niv and {D'Souza}, Richard and {Fu}, Hai and {Jones}, Amy and {Kauffmann}, Guinevere and {MacDonald}, Nicholas and {Masters}, Karen L. and {Newman}, Jeffrey A. and {Parejko}, John K. and {S{\'a}nchez-Gallego}, Jos{\'e} R. and {S{\'a}nchez}, Sebastian F. and {Schlegel}, David J. and {Thomas}, Daniel and {Wake}, David A. and {Weijmans}, Anne-Marie and {Westfall}, Kyle B. and {Zhang}, Kai},
        title = "{The Data Reduction Pipeline for the SDSS-IV MaNGA IFU Galaxy Survey}",
      journal = {\aj},
         year = 2016,
        month = oct,
       volume = {152},
       number = {4},
          eid = {83},
        pages = {83},
          doi = {10.3847/0004-6256/152/4/83},
archivePrefix = {arXiv},
       eprint = {1607.08619},
 primaryClass = {astro-ph.IM},
       adsurl = {https://ui.adsabs.harvard.edu/abs/2016AJ....152...83L}
}

@ARTICLE{MaNGA_observation_Wake_2017,
       author = {{Wake}, David A. and {Bundy}, Kevin and {Diamond-Stanic}, Aleksandar M. and {Yan}, Renbin and {Blanton}, Michael R. and {Bershady}, Matthew A. and {S{\'a}nchez-Gallego}, Jos{\'e} R. and {Drory}, Niv and {Jones}, Amy and {Kauffmann}, Guinevere and {Law}, David R. and {Li}, Cheng and {MacDonald}, Nicholas and {Masters}, Karen and {Thomas}, Daniel and {Tinker}, Jeremy and {Weijmans}, Anne-Marie and {Brownstein}, Joel R.},
        title = "{The SDSS-IV MaNGA Sample: Design, Optimization, and Usage Considerations}",
      journal = {\aj},
         year = 2017,
        month = sep,
       volume = {154},
       number = {3},
          eid = {86},
        pages = {86},
          doi = {10.3847/1538-3881/aa7ecc},
archivePrefix = {arXiv},
       eprint = {1707.02989},
 primaryClass = {astro-ph.GA},
       adsurl = {https://ui.adsabs.harvard.edu/abs/2017AJ....154...86W}
}

@ARTICLE{MaNGA_morphology_catalog_Dominguez_2017,
       author = {{Dom{\'\i}nguez S{\'a}nchez}, H. and {Margalef}, B. and {Bernardi}, M. and {Huertas-Company}, M.},
        title = "{SDSS-IV DR17: final release of MaNGA PyMorph photometric and deep-learning morphological catalogues}",
      journal = {\mnras},
         year = 2022,
        month = jan,
       volume = {509},
       number = {3},
        pages = {4024-4036},
          doi = {10.1093/mnras/stab3089},
archivePrefix = {arXiv},
       eprint = {2110.10694},
 primaryClass = {astro-ph.GA},
       adsurl = {https://ui.adsabs.harvard.edu/abs/2022MNRAS.509.4024D}
}

@ARTICLE{MaNGA_survey_Bundy_2015,
       author = {{Bundy}, Kevin and {Bershady}, Matthew A. and {Law}, David R. and {Yan}, Renbin and {Drory}, Niv and {MacDonald}, Nicholas and {Wake}, David A. and {Cherinka}, Brian and {S{\'a}nchez-Gallego}, Jos{\'e} R. and {Weijmans}, Anne-Marie and {Thomas}, Daniel and {Tremonti}, Christy and {Masters}, Karen and {Coccato}, Lodovico and {Diamond-Stanic}, Aleksandar M. and {Arag{\'o}n-Salamanca}, Alfonso and {Avila-Reese}, Vladimir and {Badenes}, Carles and {Falc{\'o}n-Barroso}, J{\'e}sus and {Belfiore}, Francesco and {Bizyaev}, Dmitry and {Blanc}, Guillermo A. and {Bland-Hawthorn}, Joss and {Blanton}, Michael R. and {Brownstein}, Joel R. and {Byler}, Nell and {Cappellari}, Michele and {Conroy}, Charlie and {Dutton}, Aaron A. and {Emsellem}, Eric and {Etherington}, James and {Frinchaboy}, Peter M. and {Fu}, Hai and {Gunn}, James E. and {Harding}, Paul and {Johnston}, Evelyn J. and {Kauffmann}, Guinevere and {Kinemuchi}, Karen and {Klaene}, Mark A. and {Knapen}, Johan H. and {Leauthaud}, Alexie and {Li}, Cheng and {Lin}, Lihwai and {Maiolino}, Roberto and {Malanushenko}, Viktor and {Malanushenko}, Elena and {Mao}, Shude and {Maraston}, Claudia and {McDermid}, Richard M. and {Merrifield}, Michael R. and {Nichol}, Robert C. and {Oravetz}, Daniel and {Pan}, Kaike and {Parejko}, John K. and {Sanchez}, Sebastian F. and {Schlegel}, David and {Simmons}, Audrey and {Steele}, Oliver and {Steinmetz}, Matthias and {Thanjavur}, Karun and {Thompson}, Benjamin A. and {Tinker}, Jeremy L. and {van den Bosch}, Remco C.~E. and {Westfall}, Kyle B. and {Wilkinson}, David and {Wright}, Shelley and {Xiao}, Ting and {Zhang}, Kai},
        title = "{Overview of the SDSS-IV MaNGA Survey: Mapping nearby Galaxies at Apache Point Observatory}",
      journal = {\apj},
         year = 2015,
        month = jan,
       volume = {798},
       number = {1},
          eid = {7},
        pages = {7},
          doi = {10.1088/0004-637X/798/1/7},
archivePrefix = {arXiv},
       eprint = {1412.1482},
 primaryClass = {astro-ph.GA},
       adsurl = {https://ui.adsabs.harvard.edu/abs/2015ApJ...798....7B}
}

@article{Martig_2009,
  author       = {Martig, Marie and Bournaud, Frédéric and Teyssier, Romain and Dekel, Avishai},
  title        = {Morphological Quenching of Star Formation: Making Early-Type Galaxies Red},
  journal      = {The Astrophysical Journal},
  volume       = {707},
  number       = {1},
  pages        = {250--267},
  year         = {2009},
  doi          = {10.1088/0004-637X/707/1/250}
}

@ARTICLE{MESA_isochrone_Choi_2016,
       author = {{Choi}, Jieun and {Dotter}, Aaron and {Conroy}, Charlie and {Cantiello}, Matteo and {Paxton}, Bill and {Johnson}, Benjamin D.},
        title = "{Mesa Isochrones and Stellar Tracks (MIST). I. Solar-scaled Models}",
      journal = {\apj},
         year = 2016,
        month = jun,
       volume = {823},
       number = {2},
          eid = {102},
        pages = {102},
          doi = {10.3847/0004-637X/823/2/102},
archivePrefix = {arXiv},
       eprint = {1604.08592},
 primaryClass = {astro-ph.SR},
       adsurl = {https://ui.adsabs.harvard.edu/abs/2016ApJ...823..102C}
}

@ARTICLE{MGE_Cappellari_2002,
       author = {{Cappellari}, Michele},
        title = "{Efficient multi-Gaussian expansion of galaxies}",
      journal = {\mnras},
         year = 2002,
        month = jun,
       volume = {333},
       number = {2},
        pages = {400-410},
          doi = {10.1046/j.1365-8711.2002.05412.x},
archivePrefix = {arXiv},
       eprint = {astro-ph/0201430},
 primaryClass = {astro-ph},
       adsurl = {https://ui.adsabs.harvard.edu/abs/2002MNRAS.333..400C}
}

@ARTICLE{Mozumdar_Knabel_2025,
       author = {{Mozumdar}, Pritom and {Knabel}, Shawn and {Treu}, Tommaso and {Sonnenfeld}, Alessandro and {Shajib}, Anowar J. and {Cappellari}, Michele and {Nipoti}, Carlo},
        title = "{XXII. Accurate stellar velocity dispersions of the SL2S lens sample and the lensing mass fundamental plane}",
      journal = {arXiv e-prints},
         year = 2025,
        month = may,
          eid = {arXiv:2505.13962},
        pages = {arXiv:2505.13962},
          doi = {10.48550/arXiv.2505.13962},
archivePrefix = {arXiv},
       eprint = {2505.13962},
 primaryClass = {astro-ph.GA},
       adsurl = {https://ui.adsabs.harvard.edu/abs/2025arXiv250513962M}
}

@MISC{MPDAF_Bacon_2016,
       author = {{Bacon}, Roland and {Piqueras}, Laure and {Conseil}, Simon and {Richard}, Johan and {Shepherd}, Martin},
        title = "{MPDAF: MUSE Python Data Analysis Framework}",
 howpublished = {Astrophysics Source Code Library, record ascl:1611.003},
         year = 2016,
        month = nov,
          eid = {ascl:1611.003},
        pages = {ascl:1611.003},
archivePrefix = {ascl},
       eprint = {1611.003},
       adsurl = {https://ui.adsabs.harvard.edu/abs/2016ascl.soft11003B}
}

@ARTICLE{Munoz_Lopez_2024,
       author = {{Mu{\~n}oz L{\'o}pez}, C. and {Krajnovi{\'c}}, D. and {Epinat}, B. and {Herrero-Alonso}, Y. and {Urrutia}, T. and {Mercier}, W. and {Bouch{\'e}}, N.~F. and {Boogaard}, L.~A. and {Contini}, T. and {Michel-Dansac}, L. and {Pessa}, I.},
        title = "{Stellar angular momentum of intermediate-redshift galaxies in MUSE surveys}",
      journal = {\aap},
         year = 2024,
        month = aug,
       volume = {688},
          eid = {A75},
        pages = {A75},
          doi = {10.1051/0004-6361/202449758},
archivePrefix = {arXiv},
       eprint = {2404.10533},
 primaryClass = {astro-ph.GA},
       adsurl = {https://ui.adsabs.harvard.edu/abs/2024A&A...688A..75M}
}

@InProceedings{MUSE_Bacon_2010,
  author        = {{Bacon}, R. and {Accardo}, M. and {Adjali}, L. and {Anwand}, H. and {Bauer}, S. and {Biswas}, I. and {Blaizot}, J. and {Boudon}, D. and {Brau-Nogue}, S. and {Brinchmann}, J. and {Caillier}, P. and {Capoani}, L. and {Carollo}, C.~M. and {Contini}, T. and {Couderc}, P. and {Daguis{\'e}}, E. and {Deiries}},
  booktitle     = {Ground-based and Airborne Instrumentation for Astronomy III},
  title         = {{The MUSE second-generation VLT instrument}},
  year          = {2010},
  editor        = {{McLean}, Ian S. and {Ramsay}, Suzanne K. and {Takami}, Hideki},
  month         = jul,
  pages         = {773508},
  series        = {Society of Photo-Optical Instrumentation Engineers (SPIE) Conference Series},
  volume        = {7735},
  adsurl        = {https://ui.adsabs.harvard.edu/abs/2010SPIE.7735E..08B},
  archiveprefix = {arXiv},
  doi           = {10.1117/12.856027},
  eid           = {773508},
  eprint        = {2211.16795},
  primaryclass  = {astro-ph.IM},
}

@ARTICLE{MUSE_pipeline_Weilbacher_2020,
       author = {{Weilbacher}, Peter M. and {Palsa}, Ralf and {Streicher}, Ole and {Bacon}, Roland and {Urrutia}, Tanya and {Wisotzki}, Lutz and {Conseil}, Simon and {Husemann}, Bernd and {Jarno}, Aur{\'e}lien and {Kelz}, Andreas and {P{\'e}contal-Rousset}, Arlette and {Richard}, Johan and {Roth}, Martin M. and {Selman}, Fernando and {Vernet}, Jo{\"e}l},
        title = "{The data processing pipeline for the MUSE instrument}",
      journal = {\aap},
         year = 2020,
        month = sep,
       volume = {641},
          eid = {A28},
        pages = {A28},
          doi = {10.1051/0004-6361/202037855},
archivePrefix = {arXiv},
       eprint = {2006.08638},
 primaryClass = {astro-ph.IM},
       adsurl = {https://ui.adsabs.harvard.edu/abs/2020A&A...641A..28W}
}

@ARTICLE{Naab_2014,
  author       = {Naab, Thorsten and Oser, Luisa and Emsellem, Eric and Cappellari, Michele and Krajnović, Davor and McDermid, Richard M. and Alatalo, Karoline and Bayet, Etienne and Blitz, Leslie and Bois, Mathieu and Bournaud, Frédéric and Bureau, Michael and Crocker, A. F. and Davies, R. L. and Davis, Timothy A. and de Zeeuw, P. Tim and Duc, P.-A. and Hirschmann, Michaela and Johansson, Peter H. and Khochfar, Sadegh and Kuntschner, Harald and Morganti, Raffaella and Oosterloo, Tom and Sarzi, Marc and Scott, Nichole and Serra, P. and van de Ven, Glenn and Weijmans, Anne-Marie and Young, L. M.},
  title        = {The {ATLAS$^{3D}$} project -- XXV. Two-dimensional kinematic analysis of simulated galaxies and the cosmological origin of fast and slow rotators},
  journal      = {Monthly Notices of the Royal Astronomical Society},
  year         = {2014},
  volume       = {444},
  number       = {4},
  pages        = {3357--3387},
  doi          = {10.1093/mnras/stt1919},
}

@ARTICLE{Naab_Ostriker_2017,
       author = {{Naab}, Thorsten and {Ostriker}, Jeremiah P.},
        title = "{Theoretical Challenges in Galaxy Formation}",
      journal = {\araa},
         year = 2017,
        month = aug,
       volume = {55},
       number = {1},
        pages = {59-109},
          doi = {10.1146/annurev-astro-081913-040019},
archivePrefix = {arXiv},
       eprint = {1612.06891},
 primaryClass = {astro-ph.GA},
       adsurl = {https://ui.adsabs.harvard.edu/abs/2017ARA&A..55...59N}
}

@article{penoyre2017,
  author = {Penoyre, Z. and Moster, B. P. and Sijacki, D. and Genel, S.},
  year = {2017},
  journal = {MNRAS},
  volume = {468},
  pages = {3883},
  title = {The angular momentum content of galaxies in the Illustris simulation: dependence on mass and morphology}
}

@ARTICLE{Rodriguez-Gomez_2017,
       author = {{Rodriguez-Gomez}, Vicente and {Sales}, Laura V. and {Genel}, Shy and {Pillepich}, Annalisa and {Zjupa}, Jolanta and {Nelson}, Dylan and {Griffen}, Brendan and {Torrey}, Paul and {Snyder}, Gregory F. and {Vogelsberger}, Mark and {Springel}, Volker and {Ma}, Chung-Pei and {Hernquist}, Lars},
        title = "{The role of mergers and halo spin in shaping galaxy morphology}",
      journal = {\mnras},
         year = 2017,
        month = may,
       volume = {467},
       number = {3},
        pages = {3083-3098},
          doi = {10.1093/mnras/stx305},
archivePrefix = {arXiv},
       eprint = {1609.09498},
 primaryClass = {astro-ph.GA},
       adsurl = {https://ui.adsabs.harvard.edu/abs/2017MNRAS.467.3083R}
}

@ARTICLE{Schulze_2018,
       author = {{Schulze}, Felix and {Remus}, Rhea-Silvia and {Dolag}, Klaus and {Burkert}, Andreas and {Emsellem}, Eric and {van de Ven}, Glenn},
        title = "{Kinematics of simulated galaxies - I. Connecting dynamical and morphological properties of early-type galaxies at different redshifts}",
      journal = {\mnras},
         year = 2018,
        month = nov,
       volume = {480},
       number = {4},
        pages = {4636-4658},
          doi = {10.1093/mnras/sty2090},
archivePrefix = {arXiv},
       eprint = {1802.01583},
 primaryClass = {astro-ph.GA},
       adsurl = {https://ui.adsabs.harvard.edu/abs/2018MNRAS.480.4636S}
}

@article{Schaye_2015_EAGLE,
  author = {Schaye, J. and Crain, R. A. and Bower, R. G. and Furlong, M. and Schaller, M. and Theuns, T. and Dalla Vecchia, C. and Frenk, C. S. and McCarthy, I. G. and Helly, J. C. and Jenkins, A. and Rosas-Guevara, Y. M. and White, S. D. M. and Baes, M. and Booth, C. M. and Camps, P. and Navarro, J. F. and Qu, Y. and Rahmati, A. and Sawala, T. and Thomas, P. A. and Trayford, J. W.},
  title = {The EAGLE project: simulating the evolution and assembly of galaxies and their environments},
  journal = {Monthly Notices of the Royal Astronomical Society},
  volume = {446},
  number = {1},
  pages = {521--554},
  year = {2015},
  doi = {10.1093/mnras/stu2058}
}

@ARTICLE{Slob_2025,
       author = {{Slob}, Martje and {Kriek}, Mariska and {de Graaff}, Anna and {Cheng}, Chloe M. and {Beverage}, Aliza G. and {Bezanson}, Rachel and {Forster Schreiber}, Natascha M. and {Lorenz}, Brian and {Mancera Pi{\~n}a}, Pavel E. and {Marchesini}, Danilo and {Muzzin}, Adam and {Newman}, Andrew B. and {Price}, Sedona H. and {Suess}, Katherine A. and {van de Sande}, Jesse and {van Dokkum}, Pieter and {Weisz}, Daniel R.},
        title = "{Fast Rotators at Cosmic Noon: Stellar Kinematics for 15 Quiescent Galaxies from JWST-SUSPENSE}",
      journal = {arXiv e-prints},
         year = 2025,
        month = jun,
          eid = {arXiv:2506.04310},
        pages = {arXiv:2506.04310},
          doi = {10.48550/arXiv.2506.04310},
archivePrefix = {arXiv},
       eprint = {2506.04310},
 primaryClass = {astro-ph.GA},
       adsurl = {https://ui.adsabs.harvard.edu/abs/2025arXiv250604310S}
}

@ARTICLE{Vande_sande_SAMI_2017,
       author = {{van de Sande}, Jesse and {Bland-Hawthorn}, Joss and {Fogarty}, Lisa M.~R. and {Cortese}, Luca and {d'Eugenio}, Francesco and {Croom}, Scott M. and {Scott}, Nicholas and {Allen}, James T. and {Brough}, Sarah and {Bryant}, Julia J. and {Cecil}, Gerald and {Colless}, Matthew and {Couch}, Warrick J. and {Davies}, Roger and {Elahi}, Pascal J. and {Foster}, Caroline and {Goldstein}, Gregory and {Goodwin}, Michael and {Groves}, Brent and {Ho}, I. -Ting and {Jeong}, Hyunjin and {Jones}, D. Heath and {Konstantopoulos}, Iraklis S. and {Lawrence}, Jon S. and {Leslie}, Sarah K. and {L{\'o}pez-S{\'a}nchez}, {\'A}ngel R. and {McDermid}, Richard M. and {McElroy}, Rebecca and {Medling}, Anne M. and {Oh}, Sree and {Owers}, Matt S. and {Richards}, Samuel N. and {Schaefer}, Adam L. and {Sharp}, Rob and {Sweet}, Sarah M. and {Taranu}, Dan and {Tonini}, Chiara and {Walcher}, C. Jakob and {Yi}, Sukyoung K.},
        title = "{The SAMI Galaxy Survey: Revisiting Galaxy Classification through High-order Stellar Kinematics}",
      journal = {\apj},
         year = 2017,
        month = jan,
       volume = {835},
       number = {1},
          eid = {104},
        pages = {104},
          doi = {10.3847/1538-4357/835/1/104},
archivePrefix = {arXiv},
       eprint = {1611.07039},
 primaryClass = {astro-ph.GA},
       adsurl = {https://ui.adsabs.harvard.edu/abs/2017ApJ...835..104V}
}

@ARTICLE{van_Dokkum_2015,
       author = {{van Dokkum}, Pieter G. and {Nelson}, Erica June and {Franx}, Marijn and {Oesch}, Pascal and {Momcheva}, Ivelina and {Brammer}, Gabriel and {F{\"o}rster Schreiber}, Natascha M. and {Skelton}, Rosalind E. and {Whitaker}, Katherine E. and {van der Wel}, Arjen and {Bezanson}, Rachel and {Fumagalli}, Mattia and {Illingworth}, Garth D. and {Kriek}, Mariska and {Leja}, Joel and {Wuyts}, Stijn},
        title = "{Forming Compact Massive Galaxies}",
      journal = {\apj},
         year = 2015,
        month = nov,
       volume = {813},
       number = {1},
          eid = {23},
        pages = {23},
          doi = {10.1088/0004-637X/813/1/23},
archivePrefix = {arXiv},
       eprint = {1506.03085},
 primaryClass = {astro-ph.GA},
       adsurl = {https://ui.adsabs.harvard.edu/abs/2015ApJ...813...23V}
}

@article{vanDokkum_1996,
  author       = {van Dokkum, P. G. and Franx, M.},
  title        = {Morphological evolution of cluster galaxies},
  journal      = {Monthly Notices of the Royal Astronomical Society},
  year         = {1996},
  volume       = {281},
  number       = {3},
  pages        = {985--1000},
  doi          = {10.1093/mnras/281.3.985}
}

@ARTICLE{Verro_2022,
       author = {{Verro}, K. and {Trager}, S.~C. and {Peletier}, R.~F. and {Lan{\c{c}}on}, A. and {Gonneau}, A. and {Vazdekis}, A. and {Prugniel}, P. and {Chen}, Y. -P. and {Coelho}, P.~R.~T. and {S{\'a}nchez-Bl{\'a}zquez}, P. and {Martins}, L. and {Arentsen}, A. and {Lyubenova}, M. and {Falc{\'o}n-Barroso}, J. and {Dries}, M.},
        title = "{The X-shooter Spectral Library (XSL): Data Release 3}",
      journal = {\aap},
         year = 2022,
        month = apr,
       volume = {660},
          eid = {A34},
        pages = {A34},
          doi = {10.1051/0004-6361/202142388},
archivePrefix = {arXiv},
       eprint = {2110.10188},
 primaryClass = {astro-ph.SR},
       adsurl = {https://ui.adsabs.harvard.edu/abs/2022A&A...660A..34V}
}

@ARTICLE{Wisnioski_2015,
       author = {{Wisnioski}, E. and {F{\"o}rster Schreiber}, N.~M. and {Wuyts}, S. and {Wuyts}, E. and {Bandara}, K. and {Wilman}, D. and {Genzel}, R. and {Bender}, R. and {Davies}, R. and {Fossati}, M. and {Lang}, P. and {Mendel}, J.~T. and {Beifiori}, A. and {Brammer}, G. and {Chan}, J. and {Fabricius}, M. and {Fudamoto}, Y. and {Kulkarni}, S. and {Kurk}, J. and {Lutz}, D. and {Nelson}, E.~J. and {Momcheva}, I. and {Rosario}, D. and {Saglia}, R. and {Seitz}, S. and {Tacconi}, L.~J. and {van Dokkum}, P.~G.},
        title = "{The KMOS$^{3D}$ Survey: Design, First Results, and the Evolution of Galaxy Kinematics from 0.7 <= z <= 2.7}",
      journal = {\apj},
         year = 2015,
        month = feb,
       volume = {799},
       number = {2},
          eid = {209},
        pages = {209},
          doi = {10.1088/0004-637X/799/2/209},
archivePrefix = {arXiv},
       eprint = {1409.6791},
 primaryClass = {astro-ph.GA},
       adsurl = {https://ui.adsabs.harvard.edu/abs/2015ApJ...799..209W}
}

@Misc{ESO2017,
  author    = {{European Southern Observatory (ESO)}},
  title     = {MUSE reduced data obtained by standard ESO pipeline processing, co-added across multiple observations},
  year      = {2017},
  copyright = {Data Access Policy for ESO Data held in the ESO Science Archive Facility},
  doi       = {10.18727/ARCHIVE/42},
  language  = {en},
  publisher = {European Southern Observatory (ESO)},
}

@Article{Oke1983,
  author   = {Oke, J.~B. and Gunn, J.~E.},
  journal  = {\apj},
  title    = {Secondary standard stars for absolute spectrophotometry.},
  year     = {1983},
  month    = mar,
  pages    = {713--717},
  volume   = {266},
  doi      = {10.1086/160817},
  url      = {https://ui.adsabs.harvard.edu/abs/1983ApJ...266..713O},
}

@Article{Salpeter1955,
  author  = {Salpeter, E. E.},
  journal = {\apj},
  title   = {The Luminosity Function and Stellar Evolution.},
  year    = {1955},
  month   = jan,
  pages   = {161--167},
  volume  = {121},
  adsurl  = {https://ui.adsabs.harvard.edu/abs/1955ApJ...121..161S},
}

@Book{VanRossum2009,
  author    = {Van Rossum, Guido and Drake, Fred L.},
  publisher = {CreateSpace},
  title     = {Python 3 Reference Manual},
  year      = {2009},
  address   = {Scotts Valley, CA},
  isbn      = {1441412697},
}

@Article{AstropyCollaboration2022,
  author        = {{Astropy Collaboration} and Price-Whelan, Adrian M. and Lim, Pey Lian and Earl, Nicholas and Starkman, Nathaniel and Bradley, Larry and Shupe, David L. and Patil, Aarya A. and Corrales, Lia and Brasseur, C.~E. and N{\"o}the, Maximilian and Donath, Axel and Tollerud, Erik and Morris, Brett M. and Ginsburg, Adam and Vaher, Eero and Weaver, Benjamin A. and Tocknell, James and Jamieson, William and van Kerkwijk, Marten H. and Robitaille, Thomas P. and Merry, Bruce and Bachetti, Matteo and G{\"u}nther, H. Moritz and Aldcroft, Thomas L. and Alvarado-Montes, Jaime A. and Archibald, Anne M. and B{\'o}di, Attila and Bapat, Shreyas and Barentsen, Geert and Baz{\'a}n, Juanjo and Biswas, Manish and Boquien, M{\'e}d{\'e}ric and Burke, D.~J. and Cara, Daria and Cara, Mihai and Conroy, Kyle E. and Conseil, Simon and Craig, Matthew W. and Cross, Robert M. and Cruz, Kelle L. and D'Eugenio, Francesco and Dencheva, Nadia and Devillepoix, Hadrien A.~R. and Dietrich, J{\"o}rg P. and Eigenbrot, Arthur Davis and Erben, Thomas and Ferreira, Leonardo and Foreman-Mackey, Daniel and Fox, Ryan and Freij, Nabil and Garg, Suyog and Geda, Robel and Glattly, Lauren and Gondhalekar, Yash and Gordon, Karl D. and Grant, David and Greenfield, Perry and Groener, Austen M. and Guest, Steve and Gurovich, Sebastian and Handberg, Rasmus and Hart, Akeem and Hatfield-Dodds, Zac and Homeier, Derek and Hosseinzadeh, Griffin and Jenness, Tim and Jones, Craig K. and Joseph, Prajwel and Kalmbach, J. Bryce and Karamehmetoglu, Emir and Ka{\l}uszy{\'n}ski, Miko{\l}aj and Kelley, Michael S.~P. and Kern, Nicholas and Kerzendorf, Wolfgang E. and Koch, Eric W. and Kulumani, Shankar and Lee, Antony and Ly, Chun and Ma, Zhiyuan and MacBride, Conor and Maljaars, Jakob M. and Muna, Demitri and Murphy, N.~A. and Norman, Henrik and O'Steen, Richard and Oman, Kyle A. and Pacifici, Camilla and Pascual, Sergio and Pascual-Granado, J. and Patil, Rohit R. and Perren, Gabriel I. and Pickering, Timothy E. and Rastogi, Tanuj and Roulston, Benjamin R. and Ryan, Daniel F. and Rykoff, Eli S. and Sabater, Jose and Sakurikar, Parikshit and Salgado, Jes{\'u}s and Sanghi, Aniket and Saunders, Nicholas and Savchenko, Volodymyr and Schwardt, Ludwig and Seifert-Eckert, Michael and Shih, Albert Y. and Jain, Anany Shrey and Shukla, Gyanendra and Sick, Jonathan and Simpson, Chris and Singanamalla, Sudheesh and Singer, Leo P. and Singhal, Jaladh and Sinha, Manodeep and Sip{\H{o}}cz, Brigitta M. and Spitler, Lee R. and Stansby, David and Streicher, Ole and {\v{S}}umak, Jani and Swinbank, John D. and Taranu, Dan S. and Tewary, Nikita and Tremblay, Grant R. and de Val-Borro, Miguel and Van Kooten, Samuel J. and Vasovi{\'c}, Zlatan and Verma, Shresth and de Miranda Cardoso, Jos{\'e} Vin{\'\i}cius and Williams, Peter K.~G. and Wilson, Tom J. and Winkel, Benjamin and Wood-Vasey, W.~M. and Xue, Rui and Yoachim, Peter and Zhang, Chen and Zonca, Andrea and {Astropy Project Contributors}},
  journal       = {\apj},
  title         = {The Astropy Project: Sustaining and Growing a Community-oriented Open-source Project and the Latest Major Release (v5.0) of the Core Package},
  year          = {2022},
  month         = aug,
  number        = {2},
  pages         = {167},
  volume        = {935},
  adsurl        = {https://ui.adsabs.harvard.edu/abs/2022ApJ...935..167A},
  archiveprefix = {arXiv},
  doi           = {10.3847/1538-4357/ac7c74},
  eid           = {167},
  eprint        = {2206.14220},
  primaryclass  = {astro-ph.IM},
}

@Article{Harris2020,
  author    = {Harris, Charles R. and Millman, K. Jarrod and van der Walt, St{\'{e}}fan J. and Gommers, Ralf and Virtanen, Pauli and Cournapeau, David and Wieser, Eric and Taylor, Julian and Berg, Sebastian and Smith, Nathaniel J. and Kern, Robert and Picus, Matti and Hoyer, Stephan and van Kerkwijk, Marten H. and Brett, Matthew and Haldane, Allan and del R{\'{\i}}o, Jaime Fern{\'{a}}ndez and Wiebe, Mark and Peterson, Pearu and G{\'{e}}rard-Marchant, Pierre and Sheppard, Kevin and Reddy, Tyler and Weckesser, Warren and Abbasi, Hameer and Gohlke, Christoph and Oliphant, Travis E.},
  journal   = {Nature},
  title     = {Array programming with {NumPy}},
  year      = {2020},
  month     = sep,
  number    = {7825},
  pages     = {357--362},
  volume    = {585},
  doi       = {10.1038/s41586-020-2649-2},
  publisher = {Springer Science and Business Media {LLC}},
}

@ARTICLE{Pulsoni_2023,
       author = {{Pulsoni}, Claudia and {Gerhard}, Ortwin and {Fall}, S. Michael and {Arnaboldi}, Magda and {Ennis}, Ana I. and {Hartke}, Johanna and {Coccato}, Lodovico and {Napolitano}, Nicola R.},
        title = "{The extended Planetary Nebula Spectrograph (ePN.S) early-type galaxy survey: The specific angular momentum of ETGs}",
      journal = {\aap},
         year = 2023,
        month = jun,
       volume = {674},
          eid = {A96},
        pages = {A96},
          doi = {10.1051/0004-6361/202346234},
archivePrefix = {arXiv},
       eprint = {2303.06132},
 primaryClass = {astro-ph.GA},
       adsurl = {https://ui.adsabs.harvard.edu/abs/2023A&A...674A..96P}
}

@ARTICLE{Pulsoni_2021,
       author = {{Pulsoni}, C. and {Gerhard}, O. and {Arnaboldi}, M. and {Pillepich}, A. and {Rodriguez-Gomez}, V. and {Nelson}, D. and {Hernquist}, L. and {Springel}, V.},
        title = "{The stellar halos of ETGs in the IllustrisTNG simulations. II. Accretion, merger history, and dark halo connection}",
      journal = {\aap},
         year = 2021,
        month = mar,
       volume = {647},
          eid = {A95},
        pages = {A95},
          doi = {10.1051/0004-6361/202039166},
archivePrefix = {arXiv},
       eprint = {2009.01823},
 primaryClass = {astro-ph.GA},
       adsurl = {https://ui.adsabs.harvard.edu/abs/2021A&A...647A..95P}
}

@ARTICLE{MAGNUS_I,
       author = {{Mozumdar}, Pritom and {Cappellari}, Michele and {Fassnacht}, Christopher D. and {Treu}, Tommaso},
        title = "{MAGNUS I: A MUSE-DEEP sample of early-type galaxies at intermediate redshift}",
      journal = {arXiv e-prints},
         year = 2025,
        month = oct,
          eid = {arXiv:2510.23863},
        pages = {arXiv:2510.23863},
          doi = {10.48550/arXiv.2510.23863},
archivePrefix = {arXiv},
       eprint = {2510.23863},
 primaryClass = {astro-ph.GA},
       adsurl = {https://ui.adsabs.harvard.edu/abs/2025arXiv251023863M}
}

@dataset{DynPop_II_dataset,
  author       = {Lu, Shengdong and
                  Zhu, Kai and
                  Cappellari, Michele and
                  Li, Ran and
                  Mao, Shude and
                  Xu, Dandan},
  title        = {The updated catalogue of "MaNGA DynPop - II.
                   Global stellar population, gradients, and star-
                   formation histories from integral-field
                   spectroscopy of 10K galaxies: link with galaxy
                   rotation, shape, and total-density gradients"
                  },
  month        = jun,
  year         = 2025,
  publisher    = {Zenodo},
  version      = 2,
  doi          = {10.5281/zenodo.15742825},
  url          = {https://doi.org/10.5281/zenodo.15742825},
}
 


\end{document}